\documentclass{egpubl}

\pdfpagewidth=\paperwidth
\pdfpageheight=\paperheight

\usepackage[T1]{fontenc}
\usepackage{arxivfonts}

\usepackage{cite}
\BibtexOrBiblatex

\printedVersion
\PrintedOrElectronic

\makeatletter
\let\ps@titlepage\ps@plain
\def\ps@preprint{%
  \let\@mkboth\@gobbletwo
  \def\@oddhead{\normalfont\footnotesize
    \hfil\@shorttitle\quad\textbullet\quad\thepage}%
  \def\@evenhead{\normalfont\footnotesize
    \thepage\quad\textbullet\quad Yusen Feng and Shangbo Luan\hfil}%
  \def\@oddfoot{\hfil\normalfont\normalsize\thepage\hfil}%
  \let\@evenfoot\@oddfoot
  \let\sectionmark\@gobble
  \let\subsectionmark\@gobble
}
\copyrightTextTitPag{}
\copyrightTextRunPag{}
\makeatother
\pdfSubject{arXiv preprint}
\hypersetup{pdfsubject={arXiv preprint},
  pdfkeywords={Material Point Method, continuum damage, viscoplasticity, reinforcement learning, robotic manipulation}}
\AddToHook{cmd/maketitle/after}{%
  \hypersetup{pdfauthor={Yusen Feng and Shangbo Luan}}%
}

\ifpdf
  \usepackage[pdftex]{graphicx}
\else
  \usepackage[dvips]{graphicx}
\fi

\usepackage{egweblnk}
\usepackage{amsmath}
\usepackage{amssymb}
\usepackage{dsfont}
\let\mathbbm\mathds
\usepackage{algorithm}
\usepackage{algpseudocode}
\usepackage{tikz}
\usetikzlibrary{positioning, arrows.meta, calc, fit}
\usepackage{xcolor}
\usepackage{xspace}
\usepackage{booktabs}
\usepackage{placeins}
\usepackage{multicol}

\makeatletter
\let\egthebibliography\thebibliography
\renewcommand{\thebibliography}[1]{%
  \egthebibliography{#1}%
  \setlength{\itemsep}{4pt}%
  \setlength{\parskip}{0pt}%
}
\makeatother

\newcommand{\ours}{{\normalfont\textsc{MG-CD-MPM}}\xspace}
\newcommand{\mpm}{{\normalfont\textsc{MPM}}\xspace}
\newcommand{\cdmpm}{{\normalfont\textsc{CD-MPM}}\xspace}
\newcommand{\mlsmpm}{{\normalfont\textsc{MLS-MPM}}\xspace}
\newcommand{\ppo}{{\normalfont\textsc{PPO}}\xspace}
\newcommand{\sac}{{\normalfont\textsc{SAC}}\xspace}

\title[Mastering Multi-State Dynamics: Continuum Damage and Viscoplastic MPM for RL-Based Dexterous Manipulation]%
      {Mastering Multi-State Dynamics: Continuum Damage and Viscoplastic MPM for RL-Based Dexterous Manipulation}

\author[Y. Feng \& S. Luan]
{\parbox{\textwidth}{\centering
Yusen Feng$^{1}$\orcid{0009-0008-8094-3610}
and Shangbo Luan$^{1}$\orcid{0009-0000-1431-7170}}
\\
{\parbox{\textwidth}{\centering
$^{1}$Peking University}}
}

\begin{document}

\teaser{%
\centering
\begin{minipage}{0.16\textwidth}\centering
\includegraphics[width=\textwidth]{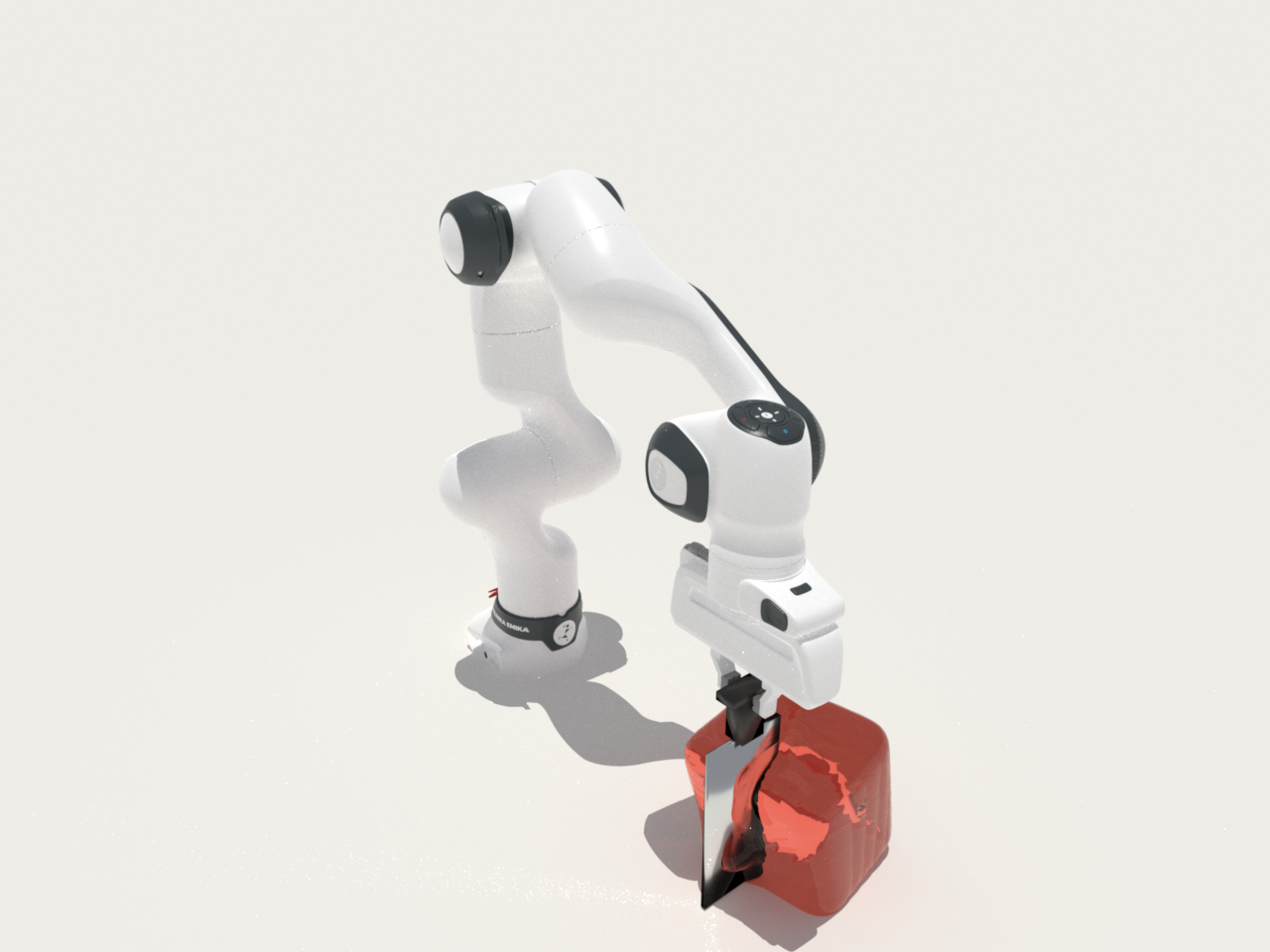}\\
{\footnotesize\normalfont Cutting jelly}
\end{minipage}\hfill
\begin{minipage}{0.16\textwidth}\centering
\includegraphics[width=\textwidth]{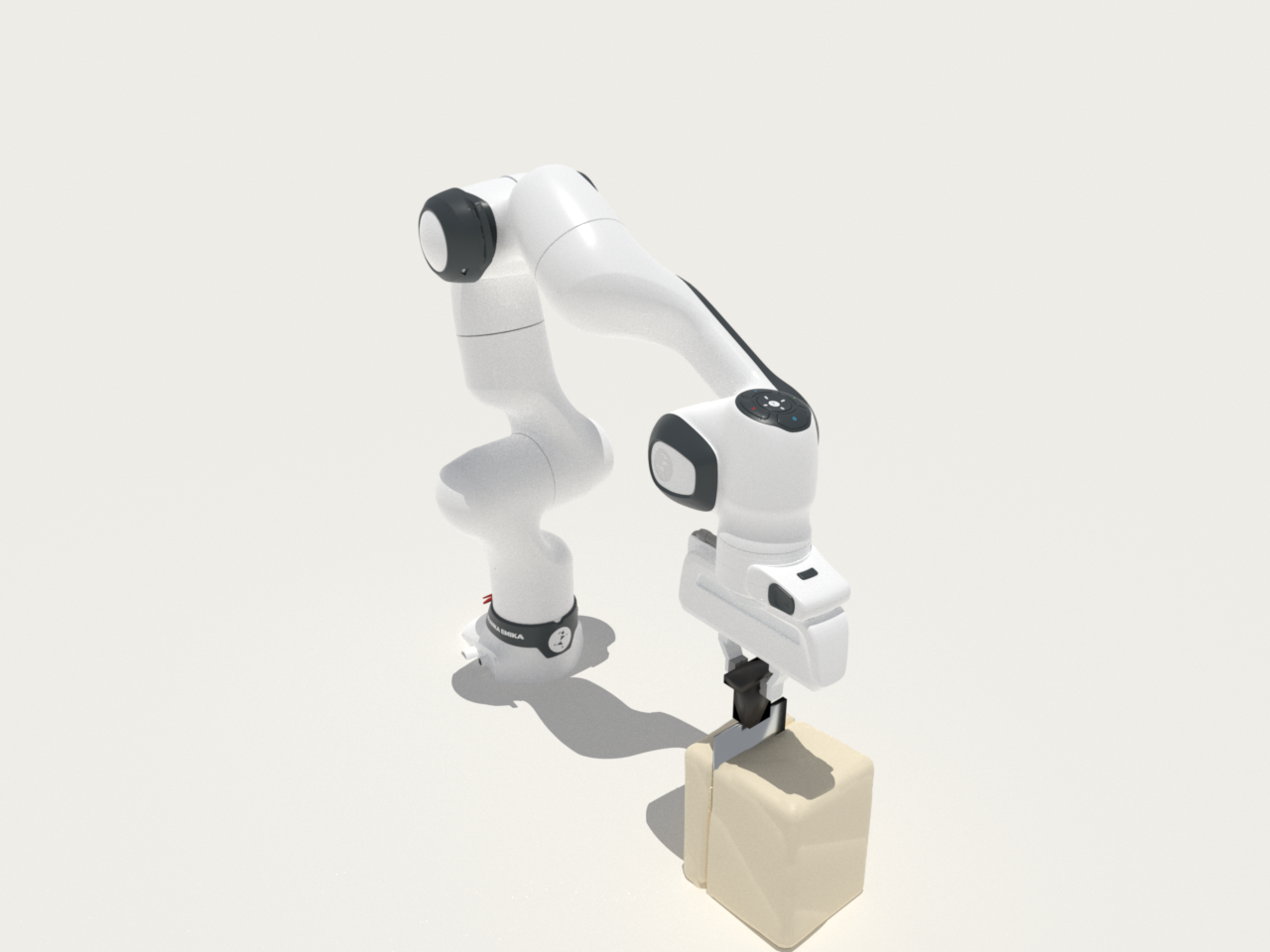}\\
{\footnotesize\normalfont Cutting tofu}
\end{minipage}\hfill
\begin{minipage}{0.16\textwidth}\centering
\includegraphics[width=\textwidth]{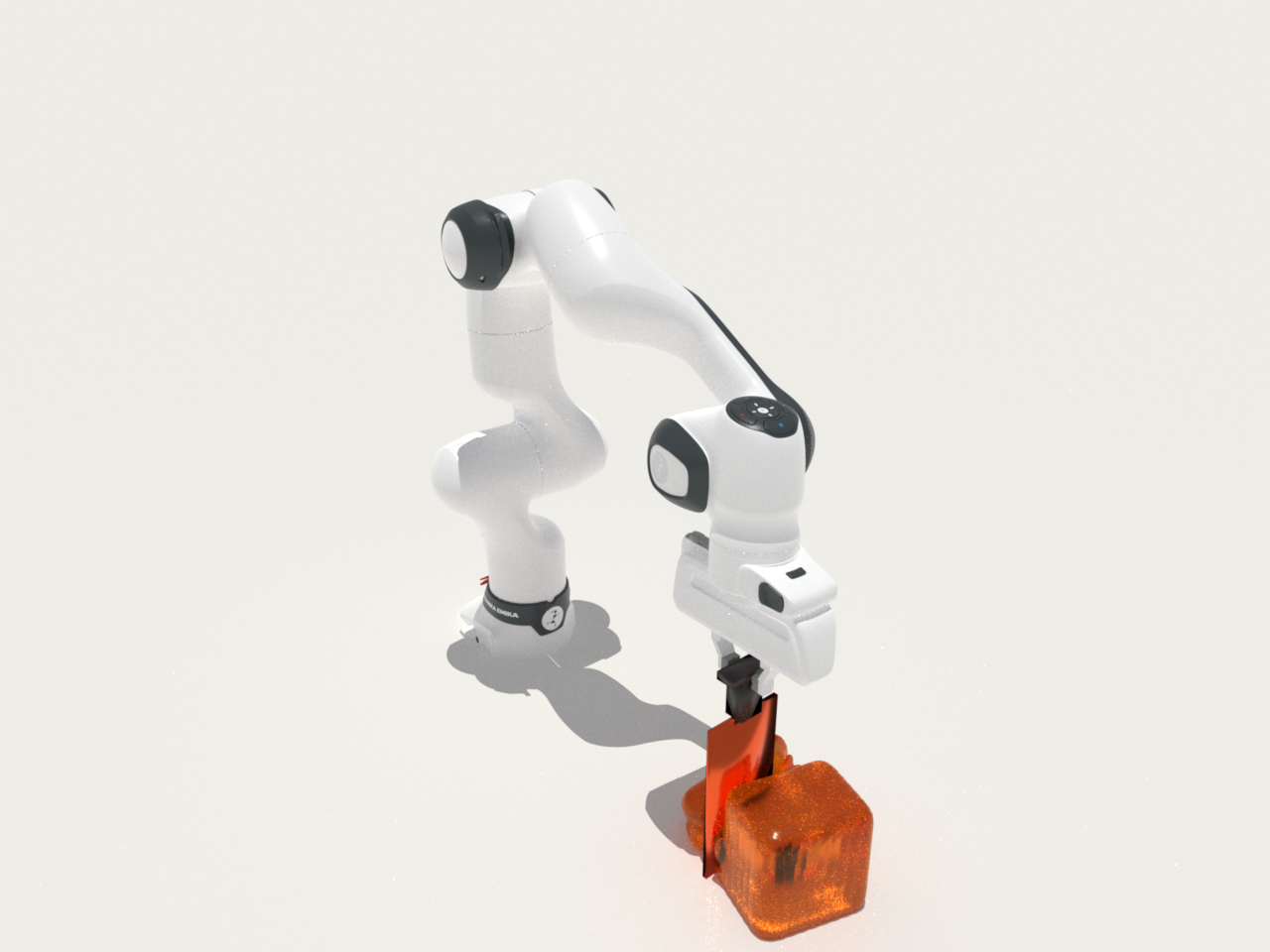}\\
{\footnotesize\normalfont Cutting wax}
\end{minipage}\hfill
\begin{minipage}{0.16\textwidth}\centering
\includegraphics[width=\textwidth]{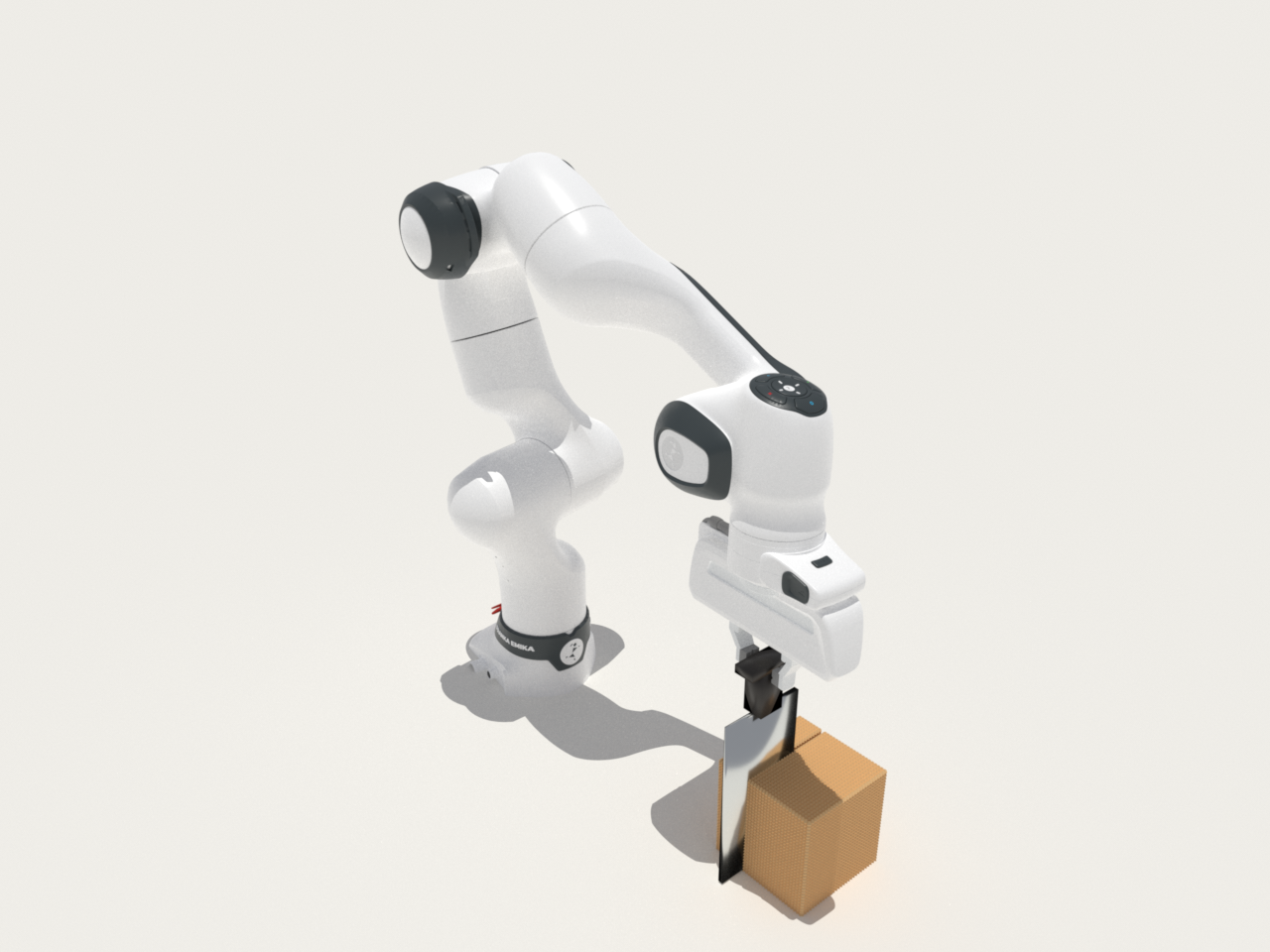}\\
{\footnotesize\normalfont Cutting peanut brittle}
\end{minipage}\hfill
\begin{minipage}{0.16\textwidth}\centering
\includegraphics[width=\textwidth]{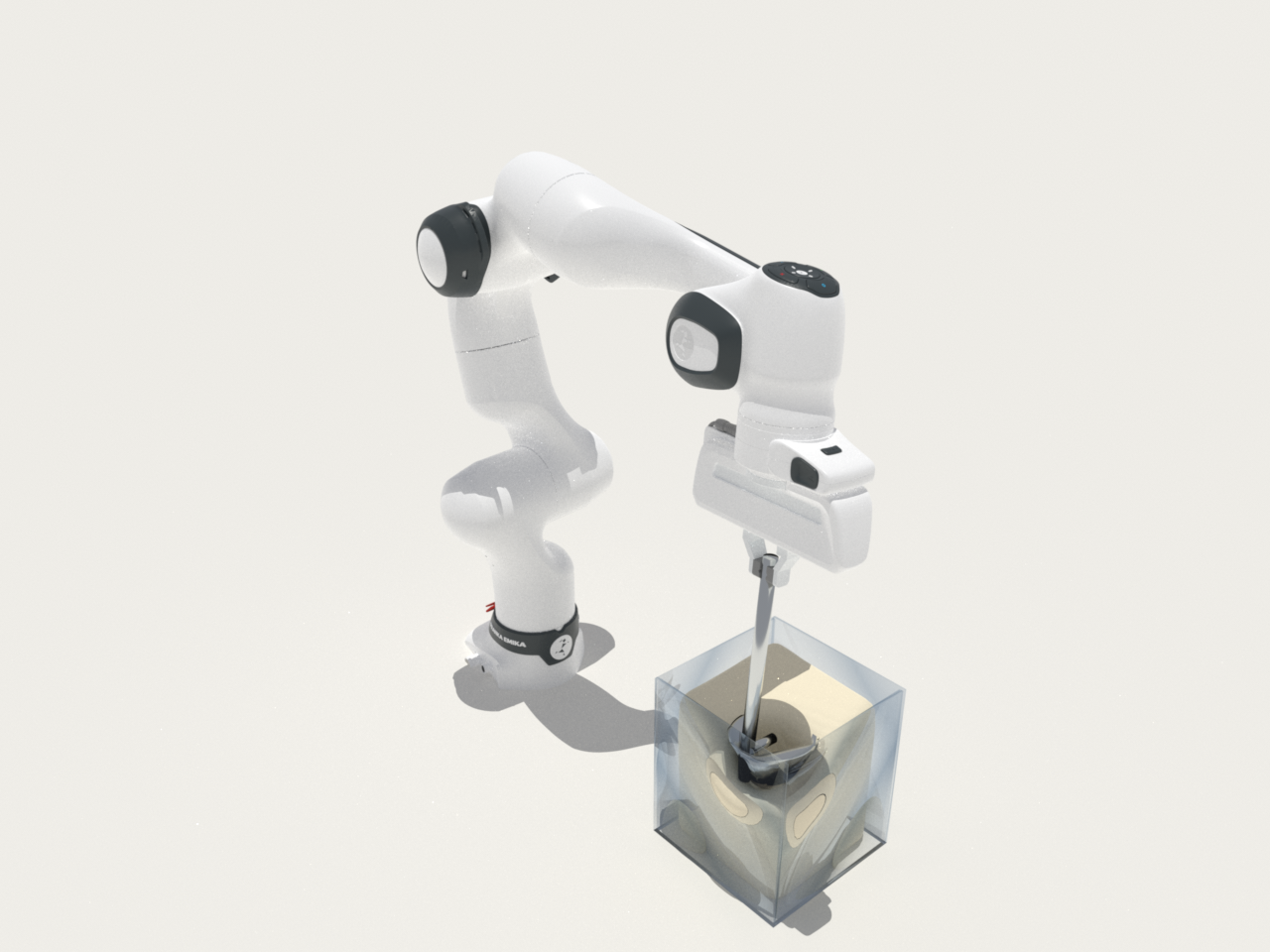}\\
{\footnotesize\normalfont Scooping ice cream}
\end{minipage}\hfill
\begin{minipage}{0.16\textwidth}\centering
\includegraphics[width=\textwidth]{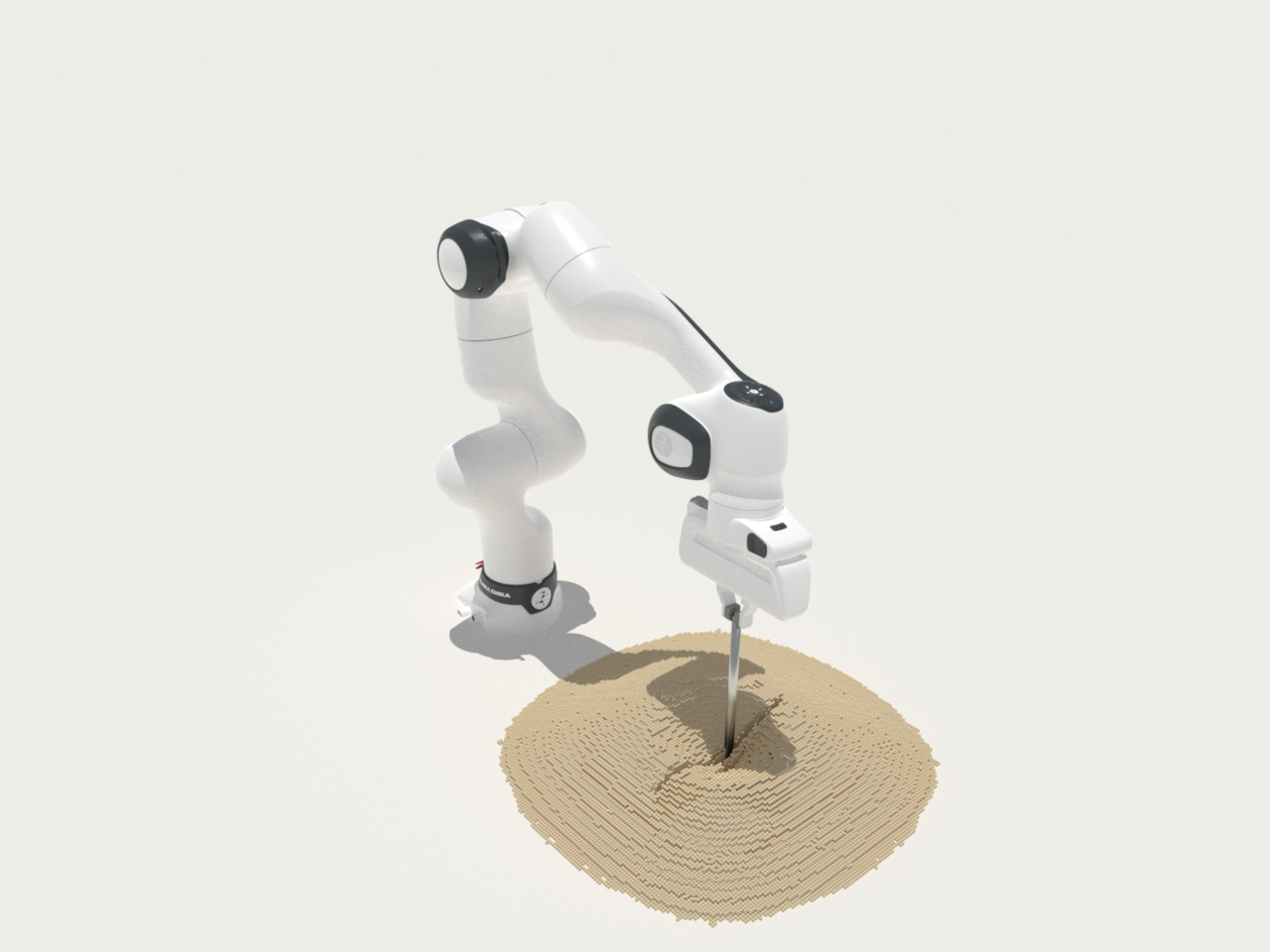}\\
{\footnotesize\normalfont Shoveling sand}
\end{minipage}
\caption{\ours{} drives six robotic manipulation scenes that span
soft, brittle, ductile, viscoplastic, and granular regimes within a
single per-particle constitutive layer. The first four scenes
demonstrate cutting across a wide stiffness and toughness range;
the last two demonstrate scooping and shoveling. Closed-loop
\ppo{} agents are trained end-to-end on the tofu and peanut brittle
cutting tasks.}
\label{fig:teaser}
}

\maketitle

\begin{abstract}
Robotic manipulation of multi-state materials needs a simulator that
handles large deformation, topology change, and two-way contact between
tools and materials while staying fast enough to drive online
reinforcement learning.
We present \ours, a tool-centered localized extension
of \cdmpm{}~\cite{WolperFangLiEtAl19}. The implicit
Ginzburg-Landau phase-field damage system is solved on a local patch by
preconditioned conjugate gradients using a geometric-multigrid V-cycle
as the preconditioner. The background \mpm{} discretization remains
uniform: only the damage solve is localized. Two-way rigid-tool coupling converts \mpm{} contact impulses into
forces and torques on the tool, exposing material resistance directly to
the policy. The resulting Gym-like simulator covers soft solids, brittle
materials, granular media, and viscoplastic or semi-fluid media within a
unified per-particle constitutive layer. We demonstrate cutting,
scooping, and shoveling in six scenes, combining scripted tool motions
with \ppo{} policies trained on two cutting tasks.

\end{abstract}

\section{Introduction}

Dexterous manipulation of multi-state materials includes cutting fruit,
scooping ice cream, shoveling sand, and splitting brittle solids. These
tasks require a simulator that handles large deformation, topology
change, and two-way contact between tools and materials while remaining
fast enough for online reinforcement learning.
The Material Point Method (\mpm{}) has become the natural particle-grid
substrate for graphics-scale soft, brittle, granular, and viscoplastic
materials \cite{StomakhinSchroederChaiEtAl13,StomakhinSchroederJiangEtAl14,KlarGastPradhanaEtAl16}. \cdmpm{}-style approaches add a damage variable so that fracture is
driven by stress rather than by mesh topology updates
\cite{WolperFangLiEtAl19,WolperChenLiEtAl20}. In parallel, robotic
cutting benchmarks have begun to target heterogeneous materials with
force feedback \cite{XuXianLinEtAl23,YangHuRavichandar26,KohSongChoiEtAl26,OuTavakoli25}.

Two practical obstacles remain when this line of work is plugged into an
RL loop. \emph{First}, \mpm{} cost grows with the number of active
particles, but direct tool contact is confined to a small region of the material: a uniform high-resolution grid spends the
rollout budget on regions the tool never reaches. UBSoft addresses this
for elastoplastic media by adapting \mpm{} resolution to gripper proximity,
but does not model damage or cutting \cite{LinFanWangEtAl24}.
\emph{Second}, \cdmpm{} resolves fracture physically through a phase-field
damage field \cite{WolperFangLiEtAl19}, but in its current form runs on
a single global grid, and is too costly to drop directly into \ppo{}-style
training that demands millions of frames.

We bridge this gap with a \cdmpm{}-style cutting simulator wrapped as a
Gym-like RL environment. The simulator drives separation through a
tool-aware damage layer, so the policy decides where cracks form rather
than replaying a kinematic blade path over a static crack model.
MGPCG solves the damage smoothing equation within a tool-centered
active region, making the implicit phase-field damage solve cost
scale with the patch volume rather than the full grid. \mpm{} contact
impulses are converted into forces and torques on the rigid tool,
allowing material resistance to feed back into policy decisions.

Concretely, our contributions are:
\begin{itemize}
\item A unified GPU-accelerated \mpm{} environment for soft solids, brittle materials,
granular media, and viscoplastic or semi-fluid media, with a shared
constitutive layer and an interface for reinforcement learning.
\item A coupled phase-field damage formulation with a local solve
around the tool, accelerated by conjugate gradients with geometric
multigrid preconditioning. The background \mpm{} grid remains uniform.
\item Bidirectional coupling between material and tool dynamics through
contact impulses, with tools held by friction in a gripper. Cutting
policies are trained with \ppo{} through rigid-body pretraining followed
by fine-tuning in the \mpm{} environment.
\end{itemize}

\section{Related Work}

\subsection{\mpm{} Simulation for Robot-Oriented Multi-State Materials}

The Material Point Method (\mpm{}) was introduced as a particle-in-cell
extension for history-dependent solid mechanics
\cite{SulskyChenSchreyer94,SulskyZhouSchreyer95} and has become a standard
graphics tool for large deformation, contact, and topology change across
snow, sand, phase-change materials, and viscoplastic media
\cite{StomakhinSchroederChaiEtAl13,StomakhinSchroederJiangEtAl14,KlarGastPradhanaEtAl16,JiangSchroederTeranEtAl16}.
Two threads in this body of work matter most for tool-driven cutting.
\mpm{} fracture methods keep a particle-grid formulation but drive
separation through damage rather than through topology updates. \cdmpm{}
evolves a Ginzburg-Landau phase-field damage variable. AnisoMPM
replaces the phase field with non-local geometric damage and structural
tensors for anisotropic materials. Ductile fracture has been formulated
through elastoplasticity with Rankine or von~Mises damage and a
particle-cloud fracture-surface visualization. A recent system pairs
an optimized \cdmpm{} with 3D Gaussian splatting for mixed-material brittle
fracture
\cite{WolperFangLiEtAl19,WolperChenLiEtAl20,WangDingGastEtAl19,HuangChenLuEtAl26}.
A parallel thread closes the loop with the tool: \mlsmpm{} with CPIC
supports cutting and two-way rigid-body coupling directly by allowing
particle-grid transfers across displacement discontinuities, building on
a longer history of coupling between rigid and deformable bodies in physics simulation
\cite{HuFangGeEtAl18,ShinarSchroederFedkiw08}.

Robot-oriented \mpm{} simulators bring this physics into specific
manipulation regimes: tactile contact, fluid manipulation, soft-robot
co-design, deformable-object reconstruction, surgical or food cutting,
and multiphysics RL stability
\cite{XianZhuXuEtAl23,WangMaSpielbergEtAl23,SiZhangBenEtAl24,ChenHuSunEtAl26,KohSongChoiEtAl26,OuTavakoli25,XingLukOh25},
and engine-level systems such as Genesis target general embodied-AI
physics \cite{GenesisAuthors24}. Closest to our scaling concern, UBSoft
drops a multi-level grid into \mlsmpm{} and adapts cell size to robot
proximity, but is restricted to elastoplastic media (sand, snow) and does
not model damage or cutting \cite{LinFanWangEtAl24}. Across this line,
no prior system combines a \cdmpm{}-style damage layer for tool-driven
separation with a tool-centered localized damage solve inside a single
uniform-grid \mlsmpm{} solver. Our simulator provides this combination and exposes it through
a Gym-like RL interface while preserving two-way rigid-tool coupling.

\subsection{Cutting Robotics: Simulators, Real Robots, and Policy Learning}

Robotic cutting combines a separation model on the simulator side with
control and feedback on the policy side. Mesh-based simulators rely on
topology updates, virtual nodes, or remeshing on tetrahedral and
polyhedral discretizations \cite{MolinoBaoFedkiw04,WuWestermannDick15}.
Among recent differentiable cutting simulators, DiSECt augments FEM with
a continuous-contact SDF and a damage-spring cohesive zone along a
prescribed cutting plane and makes the whole pipeline differentiable
\cite{HeidenMacklinNarangEtAl21}. The \mpm{} side has grown in parallel:
RoboNinja learns a tolerance-modulated cutting policy for soft objects
with rigid cores on top of \mpm{} dynamics with sparse force-collision
feedback \cite{XuXianLinEtAl23}. CRESSim-\mpm{} provides a GPU \mpm{} library
with contact between soft and rigid bodies tailored to surgical cutting and suturing
\cite{OuTavakoli25}. CulinaryCut-VLAP builds an \mlsmpm{} food-cutting
environment in which contact forces are estimated from particle-grid
impulse exchange \cite{KohSongChoiEtAl26}. TopoCut extends this line
to multi-step cutting, pairing a particle-based elastoplastic solver
with a damage-driven topology-discovery mechanism and a discrete
diffusion policy guided by spectral rewards
\cite{WangBianHeidenGarg25}.

On the policy side, model-free reinforcement learning, particularly \ppo{},
remains the practical backbone for continuous robotic
control \cite{SchulmanWolskiDhariwalEtAl17}. Broader deformable-object
benchmarks have extended learning from cloth and fluids to plasticine,
dough-like materials, dexterous deformables, and graph-based shaping
\cite{LinWangOlkinHeld21,HuangHuDuEtAl21,LiHuangChenEtAl23,ChenXuYuEtAl24,ShiXuHuangEtAl22,ShiXuClarkeEtAl23}.
Real-robot cutting work emphasizes force feedback and multi-material
structure: a recent chicken-deboning system pairs a custom multi-material
cutting simulator with a hardware testbed and uses residual reinforcement
learning on discretized force observations \cite{YangHuRavichandar26},
and the simulator side cited above is consistently motivated by
hardware-relevant force-aware behavior
\cite{XuXianLinEtAl23,KohSongChoiEtAl26,OuTavakoli25}. Across this line,
simulators are typically tuned for fracture fidelity in a
fixed-resolution scene or for a single application domain, while
real-robot pipelines often replay kinematic tool trajectories or operate
without a fast simulator behind them. Our framework closes this gap by
exposing a \cdmpm{}-style multi-material cutting model and a
tool-centered localized damage solve behind a Gym-like interface with
two-way rigid-tool feedback, so the \ppo{} policies of
Sec.~\ref{sec:experiments} train on the same contact physics the
tool will feel at deployment.

\section{Method}
\label{sec:method}

\begin{figure*}[t]
\centering
\begin{tikzpicture}[
  font=\footnotesize,
  >={Stealth[length=2mm]},
  particle/.style={rectangle, rounded corners=2pt, draw=violet!60,
      fill=violet!12, minimum height=7mm, minimum width=22mm,
      align=center, inner sep=2pt},
  gridL/.style={rectangle, rounded corners=2pt, draw=green!40!black,
      fill=green!18, minimum height=8mm, minimum width=42mm,
      align=center, inner sep=2pt},
  gridR/.style={rectangle, rounded corners=2pt, draw=teal!60!black,
      fill=teal!18, minimum height=8mm, minimum width=42mm,
      align=center, inner sep=2pt},
  toolop/.style={rectangle, rounded corners=2pt, draw=orange!85!black,
      fill=orange!18, minimum height=7mm, minimum width=32mm,
      align=center, inner sep=2pt},
  rl/.style={rectangle, rounded corners=2pt, draw=blue!50!black,
      fill=blue!10, minimum height=7mm, minimum width=32mm,
      align=center, inner sep=2pt},
  resize/.style={rectangle, rounded corners=2pt, draw=red!50!black,
      fill=red!10, minimum height=7mm, align=center, inner sep=2pt},
  flow/.style={->, thick, draw=black!65},
  bluearrow/.style={->, very thick, draw=blue!70!black},
  redarrow/.style={->, very thick, draw=red!70!black},
  orangearrow/.style={->, thick, draw=orange!85!black},
  rlarrow/.style={->, thick, draw=blue!50!black},
  dashedloop/.style={->, thick, draw=gray!60, dashed},
]
\node[particle] (v_in)  at (0,    0) {Velocity\\ $\mathbf{v}_p^n$};
\node[particle] (F_in)  at (3.2,  0) {Def.\ grad.\\ $\mathbf{F}_p^n$};
\node[particle] (d_in)  at (6.4,  0) {Phase\\ $d_p^n$};
\node[particle] (H_in)  at (9.6,  0) {History\\ $\mathcal{H}_p^n$};
\node[gridL] (p2g_m) at (1.6, -1.6) {P2G \quad (\emph{all} $p$, single grid $h_0$)};
\node[gridL] (mom)   at (1.6, -3.0) {Grid momentum $(m\mathbf{v})_i^n$};
\node[gridL] (force) at (1.6, -4.4) {Compute force (Eq.~\ref{eq:grid-force})\\ split, uses OLD $d_p^n$};
\node[gridL] (sdf)   at (1.6, -5.9) {SDF project, record $\mathbf{j}_i$};
\node[gridL] (g2p_m) at (1.6, -7.3) {G2P: $\mathbf{v}_p^{n+1}, \mathbf{C}_p^{n+1}, \mathbf{F}_p^{n+1}$};
\node[gridR] (p2g_d) at (8.0, -1.6) {P2G \quad (\emph{only} $p \in \mathcal{N}(\mathbf{T})$)};
\node[gridR] (gridd) at (8.0, -3.0) {Grid phase $d_i^n$};
\node[gridR] (solve) at (8.0, -4.4) {Phase solve (Eq.~\ref{eq:grid-d-system})\\ $\mathbf{A}_\ell \mathbf{d}^{n+1} = \mathbf{b}^n$ via MGPCG};
\node[gridR] (g2p_d) at (8.0, -7.3) {G2P + irreversibility + $\mathcal{H}$ update};
\node[resize, minimum width=98mm] (resize) at (4.8, -8.5)
  {Tool-tracking patch $\mathcal{N}(\mathbf{T})$ (Eq.~\ref{eq:patch}): damage solve runs only here; history persists outside};
\node[particle] (v_out) at (0,   -9.9) {$\mathbf{v}_p^{n+1}$};
\node[particle] (F_out) at (3.2, -9.9) {$\mathbf{F}_p^{n+1}$};
\node[particle] (d_out) at (6.4, -9.9) {$d_p^{n+1}$};
\node[particle] (H_out) at (9.6, -9.9) {$\mathcal{H}_p^{n+1}$};
\node[toolop] (wrench) at (13.6, -3.0) {Wrench $\mathbf{F}_\mathbf{T},\boldsymbol{\tau}_\mathbf{T}$\\ (Eq.~\ref{eq:wrench})};
\node[toolop] (tool)   at (13.6, -4.6) {Tool 6-DoF body};
\node[toolop] (arm)    at (13.6, -6.2) {Robot arm\\ $(\mathbf{q},\dot{\mathbf{q}})$, PD ctrl};
\node[rl]     (obs)    at (13.6, -7.8) {RL obs $\mathbf{o}^{n+1}$ (Eq.~\ref{eq:obs})};
\node[rl]     (policy) at (13.6, -9.4) {Policy $\pi_\theta \to \mathbf{a}^n$};
\draw[flow] (v_in.south)  -- (v_in.south  |- p2g_m.north);
\draw[flow] (F_in.south)  -- (F_in.south  |- p2g_m.north);
\draw[flow] (d_in.south)  -- (d_in.south  |- p2g_d.north);
\draw[flow] (p2g_m) -- (mom);
\draw[flow] (mom) -- (force);
\draw[flow] (force) -- (sdf);
\draw[flow] (sdf) -- (g2p_m);
\draw[flow] (p2g_d) -- (gridd);
\draw[flow] (gridd) -- (solve);
\draw[flow] (solve) -- (g2p_d);
\draw[flow] (g2p_m.south) -- (g2p_m.south |- resize.north);
\draw[flow] (g2p_d.south) -- (g2p_d.south |- resize.north);
\draw[flow] (v_out.north |- resize.south) -- (v_out.north);
\draw[flow] (F_out.north |- resize.south) -- (F_out.north);
\draw[flow] (d_out.north |- resize.south) -- (d_out.north);
\draw[flow] (H_out.north |- resize.south) -- (H_out.north);
\draw[bluearrow, rounded corners=3pt]
  (H_in.east) -- (11.5, 0) |- (solve.east);
\draw[redarrow, rounded corners=3pt]
  (d_in.west) -| (4.8, -4.4) -- (force.east);
\draw[orangearrow, rounded corners=3pt]
  (sdf.east) -| (11, -3.0) -- (wrench.west);
\draw[orangearrow] (wrench) -- (tool);
\draw[orangearrow] (tool) -- (arm);
\draw[rlarrow] (arm) -- (obs);
\draw[rlarrow] (obs) -- (policy);
\draw[dashedloop] (policy.east) -- ++(0.7,0) |- (arm.east);
\end{tikzpicture}
\caption{Per-step data flow of \ours at
$t^n \to t^{n+1}$, drawn in the style of \cdmpm{} Figure~13
\cite{WolperFangLiEtAl19}.
\emph{Left column} (green): momentum stream, runs on the entire grid
and uses the \emph{old} damage $d_p^n$ in the damaged-split internal
force (Eq.~\ref{eq:grid-force}).
\emph{Right column} (teal): phase-field stream, restricted to nodes
in $\mathcal{N}(\mathbf{T})$ and solved as a sparse SPD system
$\mathbf{A}_\ell \mathbf{d}^{n+1} = \mathbf{b}^n$ by MGPCG with a
geometric V-cycle preconditioner~\cite{Brandt77} (Eq.~\ref{eq:grid-d-system}).
The \textcolor{blue!70!black}{blue arrow} shows that the history
$\mathcal{H}_p^n$ derived from $\mathbf{F}_p^n$ drives the phase
solve. The \textcolor{red!70!black}{red arrow} shows that the
previous damage $d_p^n$ feeds the force computation, making our
coupling \emph{explicit-staggered}; in \cdmpm{} the red arrow points
the other way, with the force using the \emph{updated phase field}
from the phase solve. The \textcolor{orange!85!black}{orange channel} sums per-node
SDF impulses $\mathbf{j}_i$ into a rigid-tool wrench
(Eq.~\ref{eq:wrench}). It integrates the robot arm under this wrench and
closes the loop through the RL observation $\mathbf{o}^{n+1}$
(Eq.~\ref{eq:obs}) and policy $\pi_\theta$.}
\label{fig:pipeline}
\end{figure*}
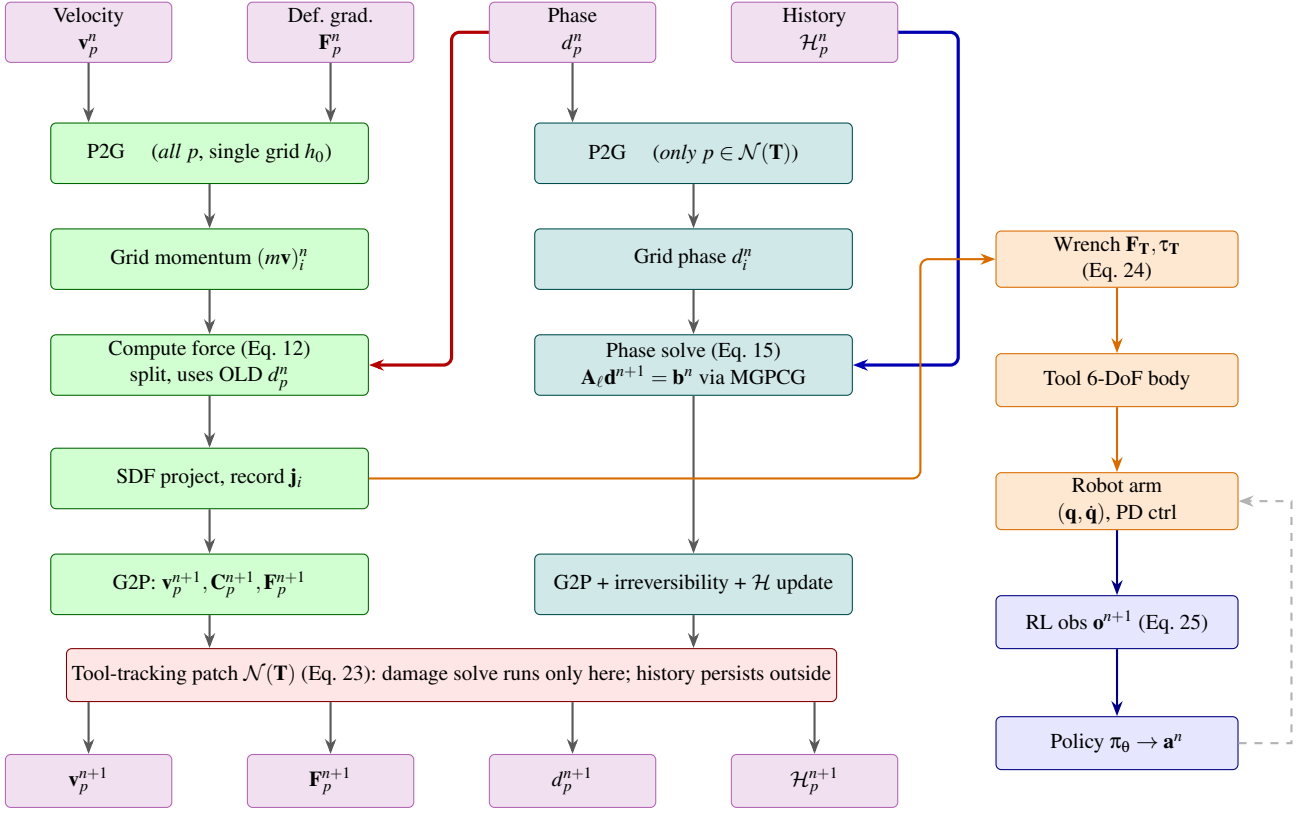

We call our method \ours. It extends the
\cdmpm{} update of~\cite{WolperFangLiEtAl19} in two ways for RL-driven
tool manipulation. First, we solve the stress-driven phase field only in
a tool-centered region $\mathcal{N}(\mathbf{T})$,
as described in Sec.~\ref{sec:patch}. Second, we precondition PCG with a 3D geometric-multigrid
V-cycle, yielding MGPCG. Crack initiation and propagation remain
stress-driven; only the solve region is localized. The same multigrid approach has
recently been applied to real-time graphics simulation in a mixed-FEM
elasticity setting~\cite{RuanWangLiuChen24}, complementary to the
\mpm{} substrate we target here. The \mpm{} background grid itself stays uniform
at cell size $h_0$ everywhere. Per timestep the simulator performs
the standard three-stage \mpm{} cycle of P2G, grid update, and
G2P, with the localized MGPCG damage solve sitting inside Stage~2; the
full step is summarized in Alg.~\ref{alg:step}. Figure~\ref{fig:pipeline}
visualizes the corresponding momentum and phase-field streams and the
tool--policy feedback loop.

\subsection{Coupled Equations of Motion}
\label{sec:eom}

Following \cdmpm{}~\cite{WolperFangLiEtAl19}, we couple momentum to a
phase-field damage variable $d \in [0,1]$, with $d = 0$ an undamaged
material point and $d = 1$ a fully cracked one. We use the damage
convention $d=1-c$, where $c$ is the intact phase field used in
\cdmpm{}.

\paragraph*{Energy split.}
A brittle solid should not crack under compression alone, so the
elastic energy is decomposed into a tensile part $\psi^+$ that drives
damage and a compressive part $\psi^-$ that does not:
\begin{equation}
\psi(\mathbf{F}) = \psi^+(\mathbf{F}) + \psi^-(\mathbf{F}),
\qquad
\mathbf{P}_d = (1-d)^2\,\mathbf{P}^{+}(\mathbf{F}) + \mathbf{P}^{-}(\mathbf{F}),
\label{eq:split}
\end{equation}
with $\mathbf{P}^\pm = \partial \psi^\pm /\partial \mathbf{F}$. The
factor $(1-d)^2$ acts only on the tensile stress $\mathbf{P}^+$. The
compressive response $\mathbf{P}^-$ remains intact even at $d=1$.
A fully cracked region therefore resists compression but has no tensile
or shear resistance. For the brittle material class
(Sec.~\ref{sec:constitutive}) we adopt the volumetric--deviatoric
split of Amor et al.~\cite{AmorMarigoMaurini09},
\begin{align}
\psi^+(\boldsymbol{\epsilon})
&= \tfrac{1}{2}\,K\,\langle\mathrm{tr}\,\boldsymbol{\epsilon}\rangle_+^2
+ \mu\,\lVert\boldsymbol{\epsilon}_{\!\text{dev}}\rVert^2,
\label{eq:psip}\\
\psi^-(\boldsymbol{\epsilon})
&= \tfrac{1}{2}\,K\,\langle\mathrm{tr}\,\boldsymbol{\epsilon}\rangle_-^2,
\label{eq:psim}
\end{align}
with bulk modulus $K$, shear modulus $\mu$, infinitesimal strain
$\boldsymbol{\epsilon} = \tfrac{1}{2}(\mathbf{F}+\mathbf{F}^\top) - \mathbf{I}$,
and its deviator $\boldsymbol{\epsilon}_{\!\text{dev}}$. The ramp
functions $\langle x\rangle_\pm = (x \pm |x|)/2$ separate the trace of
strain into volumetric expansion, which feeds $\psi^+$, and
volumetric compression, which feeds $\psi^-$. The deviator enters
$\psi^+$ in full, since shear distortion produces shear-driven
cracking regardless of the sign of the volumetric part.

\paragraph*{Evolution equations.}
The momentum and damage fields then satisfy
\begin{align}
\rho\,\dot{\mathbf{v}}
&= \nabla\!\cdot\mathbf{P}_d + \rho\,\mathbf{b},
\label{eq:momentum}\\
\eta\,\dot d
&= \underbrace{2(1-d)\,\mathcal{H}}_{\text{drive}}
\;-\;\underbrace{\frac{G_c}{\ell_c}\,d}_{\text{restoring}}
\;+\;\underbrace{G_c\,\ell_c\,\Delta d}_{\text{regularization}}.
\label{eq:damage}
\end{align}
Equation~\eqref{eq:momentum} is the standard balance of linear
momentum with body forces $\mathbf{b}$; damage enters the dynamics
only by softening $\mathbf{P}^+$ wherever $d > 0$.
Equation~\eqref{eq:damage} is the Ginzburg--Landau phase-field
equation. The drive
$2(1-d)\mathcal{H}$ increases $d$ when the past tensile energy
$\mathcal{H}$ is high, and vanishes as $d\to1$. The restoring term
$(G_c/\ell_c)d$ opposes growth and balances the drive near
$\mathcal{H}\sim G_c/(2\ell_c)$, which sets the crack-initiation
threshold. The Laplacian term $G_c\ell_c\Delta d$ spreads damage over
a band of width $\sim\ell_c$ instead of a zero-width discontinuity.
The viscosity $\eta$ stabilizes the update; we keep it small so that
$d$ follows the stress state within a substep.

\paragraph*{Driving energy and irreversibility.}
The driver $\mathcal{H}$ is the maximum past tensile elastic energy
density at each particle,
\begin{equation}
\mathcal{H}_p^{\,n}
= \max\!\bigl(\mathcal{H}_p^{\,n-1},\ \psi^+(\mathbf{F}_p^{\,n})\bigr),
\qquad \mathcal{H}_p^{\,0} = 0.
\label{eq:H}
\end{equation}
The running maximum of $\psi^+$
retains the loading history and continues to drive $d$ after the material relaxes. The damage
field itself is held to the same constraint by an explicit
projection,
\begin{equation}
d_p^{\,n+1} \;\geq\; d_p^{\,n},
\qquad \text{applied in Stage~3,}
\label{eq:irrev}
\end{equation}
which ensures that even numerical oscillations of
Eq.~\eqref{eq:damage} cannot reverse an opened crack.

\paragraph*{Tool-local activation.}
We restrict Eq.~\eqref{eq:damage} to a neighborhood of the rigid tool,
\begin{equation}
\mathcal{N}(\mathbf{T})
= \bigl\{\mathbf{x}\,:\,\phi_\mathbf{T}(\mathbf{x}) < r_d\bigr\},
\label{eq:tool-region}
\end{equation}
where $\phi_\mathbf{T}$ is the tool SDF and $r_d$ a small activation
radius (a few finest cells $h_0$ around the blade). Outside
$\mathcal{N}(\mathbf{T})$, we set $\dot d=0$ and freeze $\mathcal{H}_p$.
The particle state is then preserved: a particle that leaves with
$d_p=0.6$ keeps that value until the tool returns, while an untouched
particle remains at $d_p=0$.

The restriction in Eq.~\eqref{eq:tool-region} controls only when
new damage can grow; it does not gate force exchange. The momentum update,
damaged stress, particle--grid transfers, and rigid-tool projection still
run over the full domain. Far-field material can therefore push on the
tool, and previously damaged material transmits the stress set by its
stored damage. Section~\ref{sec:coupling} analyzes the resulting wrench.

\subsection{Three-Stage \mpm{} Update with Multigrid Damage Solve}
\label{sec:three-stage}

A timestep $t^n \to t^{n+1}$ advances particle and grid state in the
canonical P2G / grid-update / G2P sequence on a uniform background
grid of cell size $h_0$. The momentum part of the step is standard
\mlsmpm{}~\cite{HuFangGeEtAl18}; our damage solve runs in Stage~2,
where the implicit phase-field damage system is solved by MGPCG on a
tool-centered patch
(Sec.~\ref{sec:patch}).

\paragraph*{Stage 1: Particles to grid.}
Particle $p$ contributes to nodes $i$ in its quadratic-B-spline
neighborhood with weights
$w_{ip}^n = N(\mathbf{x}_i - \mathbf{x}_p^n)$. Following
\mlsmpm{}~\cite{HuFangGeEtAl18}, mass and momentum are transferred with
the per-particle affine matrix $\mathbf{C}_p^n$,
\begin{align}
m_i^n
&= \sum_p m_p\,w_{ip}^n,
\label{eq:p2g-mass}\\
(m\mathbf{v})_i^n
&= \sum_p m_p\,w_{ip}^n
\bigl[\mathbf{v}_p^n + \mathbf{C}_p^n(\mathbf{x}_i - \mathbf{x}_p^n)\bigr],
\label{eq:p2g-mom}
\end{align}
and for brittle particles in $\mathcal{N}(\mathbf{T})$ damage is
transferred mass-weighted,
\begin{equation}
d_i^n
= \frac{1}{m_i^n}\sum_{p\in\mathcal{N}(\mathbf{T})}
m_p\,w_{ip}^n\,d_p^n.
\label{eq:p2g-d}
\end{equation}

\paragraph*{Stage 2: Grid update.}
Grid momentum is integrated at every node $i$ under the internal
force from the damaged-split stress of Eq.~\eqref{eq:split} together
with external forces $\mathbf{f}_i^{\text{ext},n}$,
\begin{align}
\mathbf{f}_i^{\text{int},n}
&= -\sum_p V_p^n
\bigl[(1-d_p^n)^2\,\mathbf{P}^{+}(\mathbf{F}_p^n)
+ \mathbf{P}^{-}(\mathbf{F}_p^n)\bigr]\nabla w_{ip}^n,
\label{eq:grid-force}\\
\mathbf{v}_i^{n+1}
&= \mathbf{v}_i^n
+ \frac{\Delta t}{m_i^n}\bigl(\mathbf{f}_i^{\text{int},n}
+ \mathbf{f}_i^{\text{ext},n}\bigr).
\label{eq:grid-v}
\end{align}
Eqs.~\eqref{eq:grid-force}--\eqref{eq:grid-v} run over the entire
grid: every particle, whether inside or outside $\mathcal{N}(\mathbf{T})$,
contributes its current damaged-split stress to the internal force,
and every grid node receives a momentum update. This is what makes
far-field elastic material capable of pushing back on the tool, and
what lets a previously-damaged region transmit a stress reduced by
its frozen $d_p$.

Damage itself is advanced only on the active nodes
$i \in \mathcal{N}(\mathbf{T})$. Treating $\mathcal{H}_i^n$ and
$d_i^n$ in the source term explicitly while keeping $d_i^{n+1}$ and
the Laplacian implicit, Eq.~\eqref{eq:damage} discretizes as
\begin{equation}
\Bigl[\tfrac{\eta}{\Delta t}+\tfrac{G_c}{\ell_c}\Bigr]\,d_i^{n+1}
- G_c\,\ell_c\,(\Delta d)_i^{n+1}
= \tfrac{\eta}{\Delta t}\,d_i^n
+ 2(1-d_i^n)\,\mathcal{H}_i^n.
\label{eq:grid-d}
\end{equation}
Over the active nodes $\{i\in\mathcal{N}(\mathbf{T})\}$ this is a
sparse, symmetric positive-definite linear system
\begin{equation}
\mathbf{A}_\ell\,\mathbf{d}^{n+1}
= \mathbf{b}^{n},
\qquad
\mathbf{A}_\ell
= \Bigl[\tfrac{\eta}{\Delta t}+\tfrac{G_c}{\ell_c}\Bigr]\mathbf{I}
- G_c\,\ell_c\,\mathbf{L}_\ell,
\label{eq:grid-d-system}
\end{equation}
with $\mathbf{L}_\ell$ the standard 7-point (3D) discrete Laplacian
on the level-$\ell$ stencil of an internal multigrid hierarchy and
homogeneous-Neumann conditions at $\partial\mathcal{N}(\mathbf{T})$.
The original \cdmpm{} paper solves this implicit phase-field damage system by
Jacobi-preconditioned conjugate gradient on a single fine
grid~\cite{WolperFangLiEtAl19}. We solve
Eq.~\eqref{eq:grid-d-system} with PCG and a \emph{geometric multigrid
V-cycle} preconditioner~\cite{Brandt77}. Each cycle smooths the fine
grid, restricts the residual to coarser grids by $2^d$ full-weighting,
recurses to the coarsest level, and prolongs the correction back. We then
post-smooth on the way up. The production MGPCG solver stops when
$\lVert\mathbf r\rVert_2/\lVert\mathbf b\rVert_2<10^{-5}$ or after 20
outer iterations. Section~\ref{sec:solver-eval} compares it with
localized Jacobi-PCG and the other solver baselines. The use of multigrid as an inner solver for \mpm{}-side implicit
systems is itself well-established for implicit
time-integration~\cite{WangLiFangEtAl20}; we transplant it to the
damage-smoothing problem under the
\emph{single-resolution-grid} setting, restricted to the
tool-tracking active region of Sec.~\ref{sec:patch}. Outside
$\mathcal{N}(\mathbf{T})$ we hold $d_i^{n+1} = d_i^n$.
This is controlled localization, not an adaptive \mpm{}
discretization. Its accuracy depends on keeping the patch boundary far
enough from the evolving damage band (Sec.~\ref{sec:solver-eval}).

Independently of the damage solve, rigid-tool boundary conditions on
$\mathbf{v}_i^{n+1}$ are enforced through $\phi_\mathbf{T}$ at every
grid node $i \in \partial\mathbf{T}$ for which the SDF predicts
penetration. The per-node impulse absorbed at the boundary is given by
\begin{equation}
\mathbf{j}_i \;=\; m_i^n\,\bigl(\mathbf{v}_i^{n+1,\text{free}} - \mathbf{v}_i^{n+1}\bigr)
\label{eq:nodal-impulse}
\end{equation}
Here $\mathbf{v}_i^{n+1,\text{free}}$
is the free velocity from Eq.~\eqref{eq:grid-v} before projection. The impulse
is the momentum the tool extracts from the material at that node. It
feeds the wrench accumulation of Sec.~\ref{sec:coupling}, providing
the policy with its force-feedback signal whether or not the
contacting node lies inside $\mathcal{N}(\mathbf{T})$.

\paragraph*{Stage 3: Grid to particles.}
Particles gather velocity, affine state, and deformation gradient. Particles satisfying $p \in \mathcal{N}(\mathbf{T})$ also gather damage:
\begin{align}
\mathbf{v}_p^{n+1}
&= \sum_i w_{ip}^n\,\mathbf{v}_i^{n+1},
\label{eq:g2p-v}\\
\mathbf{C}_p^{n+1}
&= \frac{4}{h_0^2}\sum_i w_{ip}^n\,
\mathbf{v}_i^{n+1}\,(\mathbf{x}_i - \mathbf{x}_p^n)^{\!\top},
\label{eq:g2p-C}\\
\mathbf{F}_p^{n+1}
&= \bigl(\mathbf{I} + \Delta t\,\mathbf{C}_p^{n+1}\bigr)\,\mathbf{F}_p^n,
\label{eq:g2p-F}\\
d_p^{n+1}
&= \sum_i w_{ip}^n\,d_i^{n+1},
\qquad p \in \mathcal{N}(\mathbf{T}),
\label{eq:g2p-d}\\
\mathbf{x}_p^{n+1}
&= \mathbf{x}_p^n + \Delta t\,\mathbf{v}_p^{n+1}.
\label{eq:g2p-x}
\end{align}
We then enforce the irreversibility constraint of Eq.~\eqref{eq:irrev}
as a per-particle projection,
\begin{equation}
d_p^{n+1} \;\leftarrow\;
\max\!\bigl(d_p^{n+1},\ d_p^{\,n}\bigr),
\label{eq:irrev-proj}
\end{equation}
and refresh the history variable $\mathcal{H}_p$ via Eq.~\eqref{eq:H}
using the updated deformation gradient $\mathbf{F}_p^{n+1}$. When
$d_p^{n+1}$ crosses the separation threshold $d^\ast$, particle $p$
is tagged \emph{separated}, and subsequent P2G transfers across its
cluster boundary use the CPIC discontinuity machinery
of~\cite{HuFangGeEtAl18}.

\begin{algorithm}[t]
\caption{One simulation step at time $t^n \to t^{n+1}$.}
\label{alg:step}
\begin{algorithmic}[1]
\Statex \emph{Stage 1: Particles to grid}
\For{each particle $p$ scattering to nodes $i$ on the background grid}
  \State scatter mass and affine momentum
  \Statex \hfill (Eqs.~\ref{eq:p2g-mass},~\ref{eq:p2g-mom})
  \If{$p \in \mathcal{N}(\mathbf{T})$}
    \State scatter damage to $d_i^n$ \hfill (Eq.~\ref{eq:p2g-d})
  \EndIf
\EndFor
\Statex \emph{Stage 2: Grid update}
\For{each grid node $i$}
  \State assemble $\mathbf{f}_i^{\text{int},n}$ from split damaged stress
  \Statex \hfill (Eq.~\ref{eq:grid-force})
  \State update $\mathbf{v}_i^{n+1}$ \hfill (Eq.~\ref{eq:grid-v})
  \State enforce rigid-tool SDF boundary, record impulse $\mathbf{j}_i$
\EndFor
\State solve sparse SPD system $\mathbf{A}_\ell\,\mathbf{d}^{n+1}{=}\mathbf{b}^n$ on $\mathcal{N}(\mathbf{T})$ via MGPCG
\Statex \hfill (Eq.~\ref{eq:grid-d-system})
\Statex \emph{Stage 3: Grid to particles, then resize}
\For{each particle $p$}
  \State gather $\mathbf{v}_p^{n+1}, \mathbf{C}_p^{n+1}, \mathbf{F}_p^{n+1}, \mathbf{x}_p^{n+1}$
  \Statex \hfill (Eqs.~\ref{eq:g2p-v}--\ref{eq:g2p-x})
  \If{$p \in \mathcal{N}(\mathbf{T})$}
    \State gather $d_p^{n+1}$ from grid \hfill (Eq.~\ref{eq:g2p-d})
    \State irreversibility: $d_p^{n+1}\!\leftarrow\!\max(d_p^{n+1},d_p^n)$
    \Statex \hfill (Eq.~\ref{eq:irrev-proj})
    \State update history $\mathcal{H}_p^{n+1}{=}\max(\mathcal{H}_p^n,\psi^+(\mathbf{F}_p^{n+1}))$
    \Statex \hfill (Eq.~\ref{eq:H})
    \State tag separated if $d_p^{n+1}>d^\ast$
  \EndIf
\EndFor
\Statex \emph{Two-way tool integration}
\State accumulate $(\mathbf{F}_\mathbf{T}, \boldsymbol{\tau}_\mathbf{T})$
from $\{\mathbf{j}_i\}$ \hfill (Eq.~\ref{eq:wrench})
\State integrate tool 6-DoF state under wrench plus policy action $\mathbf{a}^n$
\end{algorithmic}
\end{algorithm}

\subsection{Tool-Tracking Active Solve Region}
\label{sec:patch}

We exploit a structural property of tool-driven cutting: at any
given moment the damage field is non-trivial only in a small
neighborhood of the tool tip; far from the tool the material is
intact and the damage scalar stays at its equilibrium value
$c \equiv 1$ (equivalently $d \equiv 0$). In such an intact region,
the implicit phase-field system of Eq.~\eqref{eq:grid-d-system} has a
zero damage state and no active source,
so a full-grid V-cycle would spend most of its work in cells where
it has nothing to update.

We therefore restrict the damage solve to an axis-aligned patch centered
on the tool,
\begin{equation}
\mathcal{N}(\mathbf{T})
\;=\; \bigl\{\,(i,j,k)\in\mathrm{grid}\ :\
\lvert i-i_\mathbf{T}\rvert < h_x,\
\lvert j-j_\mathbf{T}\rvert < h_y,\
\lvert k-k_\mathbf{T}\rvert < h_z\,\bigr\},
\label{eq:patch}
\end{equation}
with $(i_\mathbf{T}, j_\mathbf{T}, k_\mathbf{T})
= \lfloor (\mathbf{x}_\mathbf{T} - \mathbf{x}_0)/h_0\rfloor$ the
integer cell index of the tool tip, and $(h_x, h_y, h_z)$ half-extents
chosen so that the regularized damage band of width $\sim \ell_c$
never reaches $\partial\mathcal{N}(\mathbf{T})$. In practice, the damage
field is smoothed every $k$ \mpm{} substeps. We
(i)~copy the patch $\mathbf{b}\leftarrow d_{\text{grid}}[\mathcal{N}(\mathbf{T})]$,
(ii)~run residual-controlled MGPCG on the patch with
homogeneous Neumann boundary conditions, and (iii)~write back
$d_{\text{grid}}[\mathcal{N}(\mathbf{T})]\leftarrow$ MGPCG output.
Outside $\mathcal{N}(\mathbf{T})$, we keep the historical field
unchanged. Nodes that enter the patch later are initialized from this
stored field. Localization is accurate when the patch boundary stays far
from the evolving damage band; Sec.~\ref{sec:solver-eval} measures the
resulting error as the patch size varies. The damage-smoothing
cost becomes proportional to $\lvert\mathcal{N}(\mathbf{T})\rvert$
rather than to the full grid, an order-of-magnitude saving when the
tool occupies a small fraction of the scene, which is the operating
regime of every task in Sec.~\ref{sec:experiments}.

\begin{figure}[t]
\centering
\begin{tikzpicture}[
  font=\footnotesize,
  >={Stealth[length=1.5mm]},
]
\fill[blue!8] (-2.4,-0.1) rectangle (2.4,2.6);
\foreach \x in {-2.4,-2.0,...,2.4}
    \draw[blue!18,very thin] (\x,-0.1) -- (\x,2.6);
\foreach \y in {-0.1,0.3,...,2.6}
    \draw[blue!18,very thin] (-2.4,\y) -- (2.4,\y);
\fill[blue!15] (-1.8,0.0) rectangle (1.8,2.0);
\draw[blue!35!black,thick] (-1.8,0.0) rectangle (1.8,2.0);
\shade[left color=red!70!black, right color=red!10]
    (-0.42,0.4) rectangle (-0.18,2.0);
\shade[left color=red!70!black, right color=red!10]
    (0.78,0.6) rectangle (1.02,2.0);
\fill[gray!40] (-0.34,1.6) rectangle (-0.26,2.6);
\node[gray!60!black,anchor=north,font=\scriptsize] at (-0.30,0.4) {$\mathbf{T}^{n-k}$};
\fill[black] (0.86,1.4) rectangle (0.94,2.6);
\node[anchor=north,font=\scriptsize] at (0.90,0.6) {$\mathbf{T}^n$};
\draw[orange!50,dashed,thick] (-0.80,0.20) rectangle (0.20,2.55);
\draw[orange!85!black,dashed,very thick]
    (0.34,0.20) rectangle (1.50,2.55);
\node[orange!70!black,anchor=south west,font=\scriptsize]
    at (0.34,2.58) {$\mathcal{N}(\mathbf{T}^n)$};
\node[gray,anchor=south east,font=\scriptsize]
    at (0.20,2.58) {$\mathcal{N}(\mathbf{T}^{n-k})$};
\draw[->, thick, orange!85!black]
    (-0.30,2.78) .. controls (0.30,3.10) .. (0.90,2.78);
\node[orange!70!black,font=\scriptsize,anchor=south]
    at (0.30,3.05) {patch tracks tool};
\node[font=\scriptsize, align=center]
    at (1.55,1.2) {\textbf{inside}\\ MGPCG\\ damage solve};
\draw[->, very thin] (1.55,0.85) -- (1.18,0.8);
\node[blue!50!black, font=\scriptsize, align=center]
    at (-1.95,1.0) {\textbf{outside}\\ identity\\ $c\equiv 1$};
\draw[->, very thin, blue!50!black] (-1.95,0.55) -- (-1.55,0.55);
\end{tikzpicture}
\caption{The tool-tracking active region
$\mathcal{N}(\mathbf{T})$ of Eq.~\eqref{eq:patch}. The
background \mlsmpm{} grid (light blue) is uniform at cell size
$h_0$ everywhere; only the MGPCG phase-field damage solve is
restricted to a half-extent
$(h_x, h_y, h_z)$ axis-aligned box centered on the tool
(dashed orange). The box rigidly tracks the tool tip from
step $n-k$ to step $n$. Inside the box the implicit phase-field
solve of Eq.~\eqref{eq:grid-d-system} sees the developing
damage band (red gradient); outside the box the material is
intact ($c\equiv 1$) so $\mathbf{A}_\ell\,d=d$ holds exactly and
no work is required. The \mpm{} transfer and constitutive update of
Sec.~\ref{sec:three-stage} are unchanged on the full grid.}
\label{fig:active-patch}
\end{figure}
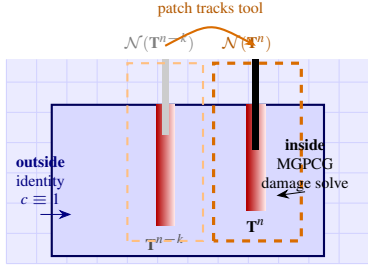

Figure~\ref{fig:active-patch} illustrates how the active damage-solve
region follows the tool while the background \mpm{} discretization remains
uniform.

\paragraph*{Solver localization.}
Eq.~\eqref{eq:patch} adapts the \emph{solver},
not the \emph{discretization}. The \mpm{} background grid stays uniform
at cell size $h_0$ everywhere; P2G and G2P transfers of
Sec.~\ref{sec:three-stage} are unchanged. We do \emph{not} subdivide
cells near the tool, do not maintain a sparse hierarchical grid in
the sense of UBSoft~\cite{LinFanWangEtAl24}, and do not perform
particle split / merge across resolution levels. The ``hierarchy''
we exploit is purely internal to the V-cycle preconditioner used by
PCG: each preconditioner application recursively
visits a ladder of progressively coarser copies of
$\mathcal{N}(\mathbf{T})$ at cell sizes $h_\ell = 2^\ell h_0$,
$\ell = 0, \dots, L$, applying a Jacobi smoother on each level and
prolonging the correction back to the finest patch. Long-wavelength
damage error that Jacobi does not efficiently remove on the finest grid becomes
short-wavelength error on a coarser one where Jacobi removes it in a
handful of iterations.

\subsection{Material Constitutive Layer}
\label{sec:constitutive}

A per-particle tag
$m_p \in \{\mathrm{S}, \mathrm{B}, \mathrm{G}, \mathrm{V}\}$ selects
how $\mathbf{P}(\mathbf{F}_p)$ in Eq.~\eqref{eq:momentum} and the
optional return mapping are evaluated:
\begin{itemize}
\item \textbf{Soft elastic-plastic solids} (S): fixed-corotated
elasticity with an optional von~Mises return mapping; damage is
disabled ($d_p \equiv 0$).
\item \textbf{Brittle solids} (B): linear elasticity with the
volumetric--deviatoric energy split of
Eqs.~\eqref{eq:psip}--\eqref{eq:psim}, coupled to the phase-field
damage of Eq.~\eqref{eq:damage}.
\item \textbf{Granular media} (G): Drucker--Prager elastoplasticity in
the formulation of~\cite{KlarGastPradhanaEtAl16}; $d_p \equiv 0$.
\item \textbf{Viscoplastic / semi-fluid materials} (V):
Herschel--Bulkley with apparent viscosity
$\eta_{\mathrm{HB}}(\dot\gamma)$; $d_p \equiv 0$.
\end{itemize}
The tag dispatches to the right routine in a single pass over
particles, so a single rollout can mix all four classes.

\subsection{Two-Way Rigid-Tool Coupling}
\label{sec:coupling}

The rigid tool $\mathbf{T}$ is a 6-DoF body with mass $M$ and
inertia tensor $\mathbf{I}$. Its SDF $\phi_\mathbf{T}$ is attached to the
tool frame. Stage~2 records the boundary impulses $\mathbf{j}_i$
(Eq.~\eqref{eq:nodal-impulse}). We combine them with the node positions
$\mathbf{x}_i$ to form a wrench at the tool's center of mass:
\begin{equation}
\mathbf{F}_\mathbf{T}
= \sum_{i \in \partial\mathbf{T}} \mathbf{j}_i,
\qquad
\boldsymbol{\tau}_\mathbf{T}
= \sum_{i \in \partial\mathbf{T}}
(\mathbf{x}_i - \mathbf{x}_\mathbf{T}) \times \mathbf{j}_i.
\label{eq:wrench}
\end{equation}
The sums in Eq.~\eqref{eq:wrench} include every SDF-boundary
node, whether or not it lies in $\mathcal{N}(\mathbf{T})$. The wrench is
therefore the full mechanical contact, not a damage-gated signal. We
integrate the tool state under this wrench and the policy action. Its
motion thus reflects material resistance rather than a replayed kinematic
trajectory.

\paragraph*{What damage does to the wrench.}
Damage affects the wrench through the softened stress
$\mathbf{P}_d$, which determines the impulses $\mathbf{j}_i$ in Stage~2
(Eq.~\ref{eq:grid-force}). The policy sees three phases during a cut.
\emph{(i)~Approach.} Near intact material ($d_p\approx0$), the tool feels
the full stiffness $\mathbf{P}^++\mathbf{P}^-$, so
$\lVert\mathbf{F}_\mathbf{T}\rVert$ rises with engagement depth.
\emph{(ii)~Damage growth.} The MGPCG damage solve grows $d_p$ in
$\mathcal{N}(\mathbf{T})$. The factor $(1-d_p)^2$ reduces $\mathbf{P}^+$,
while $\mathbf{P}^-$ preserves compressive support. \emph{(iii)~Separation.}
When $d_p$ reaches $d^\ast$, CPIC transfers~\cite{HuFangGeEtAl18} stop momentum
sharing across the cut, and the wrench drops as the blade clears. This
force profile distinguishes a clean cut from stalled penetration and is
available to the policy through its force observations
(Sec.~\ref{sec:interface}).

\subsection{Gym Interface and Policy Backbones}
\label{sec:interface}

The simulator provides the standard Gym methods \texttt{step(action)} and
\texttt{reset()}. The robot is a 7-DoF FR3 arm;
the cutting / scooping / shoveling tool is
\emph{held by friction} in the parallel gripper rather than rigidly
welded to the end-effector. The policy must therefore both execute
the manipulation and maintain enough grip force to keep the tool
from slipping under reactive contact wrench.

\paragraph*{Observation.}
The observation
$\mathbf{o}_t \in \mathbb{R}^{21+n_{\mathrm{task}}}$ concatenates
joint state and a coarse tool/scene summary:
\begin{equation}
\mathbf{o}_t = \bigl(\,
\mathbf{q}_t,\;
\dot{\mathbf{q}}_t,\;
\mathbf{x}_\mathbf{T}^{\mathrm{tip}},\;
\mathbf{q}_\mathbf{T}^{\mathrm{tip}},\;
\mathbf{e}_t^{\mathrm{task}}
\,\bigr),
\label{eq:obs}
\end{equation}
The vector contains the arm position $\mathbf{q}_t$ and velocity
$\dot{\mathbf{q}}_t$, the tool-tip position $\mathbf{x}_\mathbf{T}^{\mathrm{tip}}$
and orientation $\mathbf{q}_\mathbf{T}^{\mathrm{tip}}$, plus a small
task summary $\mathbf{e}_t^{\mathrm{task}}$. The task summary usually contains at most four scalars,
such as alive-particle ratio, tip-to-target distance, and tip height.
The wrench $(\mathbf{F}_\mathbf{T},\boldsymbol{\tau}_\mathbf{T})$ is not
given to the policy. It enters the rigid-body dynamics and deflects the
arm from its commanded targets; the resulting velocity appears in the
next observation. \ppo{} normalizes $\mathbf{o}_t$ with running mean and
variance estimates.

\paragraph*{Action.}
The action $\mathbf{a}_t\in[-1,1]^8$ concatenates a normalized
7-DoF FR3 joint-position delta around a fixed home pose with a 1-DoF
gripper-opening command:
\begin{equation}
\mathbf{a}_t \;=\; (\mathbf{a}_t^{\mathrm{arm}},\;a_t^{\mathrm{grip}}),
\qquad
\begin{aligned}
\mathbf{q}_t^{\mathrm{tgt}}
&\;=\; \mathbf{q}_{\mathrm{home}}
+ \mathbf{a}_t^{\mathrm{arm}} \odot \mathbf{r}, \\
w_t^{\mathrm{tgt}}
&\;=\; \tfrac{w_{\max}}{2}\bigl(1 + a_t^{\mathrm{grip}}\bigr),
\end{aligned}
\label{eq:action}
\end{equation}
Here $\odot$ is elementwise multiplication and
$\mathbf{r}=(2.5,1.7,2.5,1.7,2.5,2.0,2.5)\,\mathrm{rad}$ is the FR3
joint range. $w_{\max}$ is the commanded finger separation. Joint and
gripper targets drive stiff PD controllers at $50\,\mathrm{Hz}$, while
the \mpm{} update uses multiple substeps for stability. Because the tool
is held by friction, the policy must coordinate arm motion and grip
closure; if the reactive wrench overcomes friction, the tool slips and
the cut is lost.

\paragraph*{Reward.}
Each task defines its own reward as a weighted sum of interpretable
shaping terms computed from the \mpm{} field and rigid-body state.
Common ingredients are tip-to-target proximity, depth progress
into the material, blade-axis alignment, a damage-driven cut-quality
term that rewards a clean separation band, and joint-effort and
tool-slip regularizers; the per-task formulas and weights are
given in Sec.~\ref{sec:rewards}.

\paragraph*{Policy backbone: PPO with optional BC warm-start.}
We use \ppo{}~\cite{SchulmanWolskiDhariwalEtAl17} as the primary
RL backbone, optionally warm-started from a scripted action prior. \ppo{}
is on-policy: each update uses a fresh rollout, so the critic and advantage
depend on the latest batch. On the tofu cutting task, a randomly initialized
single-env actor usually stays near the ``do-nothing'' return of $-27$ to
$+27$; random exploration rarely forms a clean sequence of descent, engagement, and cutting. Two remedies help. A 64-env rollout averages gradient noise and
reaches mean return $\sim\!+328$, removing $\sim\!2600$ particles per
episode. A BC warm start initializes the actor from a scripted action prior: the
scripted action reaches $\sim\!+325$, and subsequent single-env \ppo{}
refinement holds a deterministic return of $\sim\!+278$.
Hyperparameters are given in Sec.~\ref{sec:implementation}.

\section{Experiments}
\label{sec:experiments}

\begin{figure}[t]
\centering
\includegraphics[width=\linewidth]{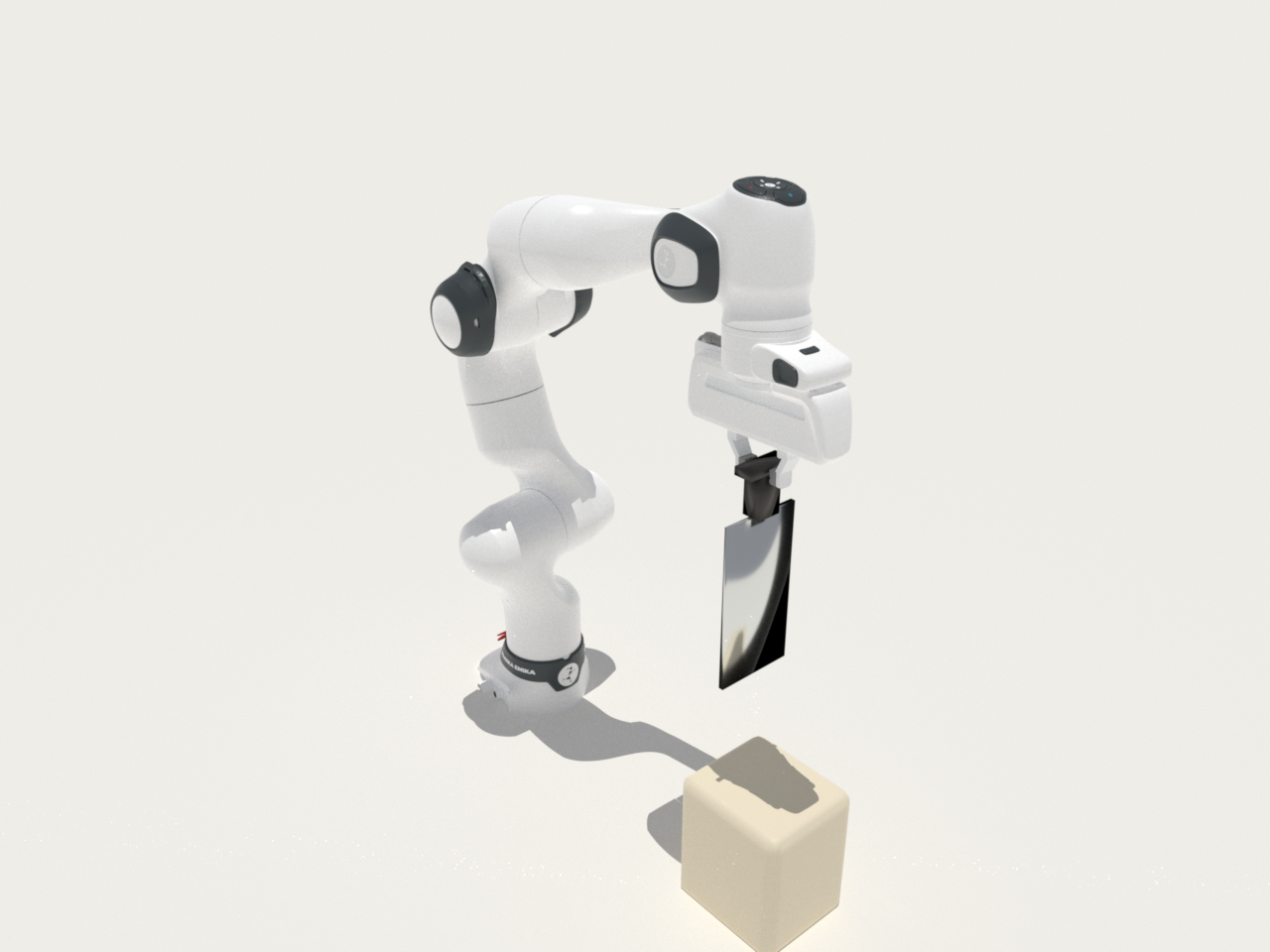}
\caption{Experimental setting for the tofu cutting task (Mitsuba~3 path-traced render of the initial
state). A 7-DoF FR3 arm grips a kitchen knife by
friction in its parallel-jaw gripper without a rigid weld and faces
a tofu block resting on a flat table. The policy commands
incremental joint targets every $20\,\mathrm{ms}$ ($50\,\mathrm{Hz}$
control rate); contact forces from the \ours{} damage solve are
fed back to the arm and gripper through the two-way coupling of
Sec.~\ref{sec:coupling}, so a wrench large enough to overcome
finger friction will physically dislodge the knife and the cut is
lost. The peanut brittle cutting task reuses the same scene with a stiffer,
higher-toughness slab.}
\label{fig:tofu-setting}
\end{figure}

Figure~\ref{fig:tofu-setting} shows the shared robot--tool arrangement
used by the two closed-loop cutting case studies.

We evaluate the simulator on six tool-manipulation demos
(Figure~\ref{fig:teaser}) that together span the multi-state regime
introduced in Sec.~\ref{sec:constitutive}: four cutting tasks
across a soft-to-brittle stiffness/toughness spectrum, plus a
viscoplastic scooping scene and a granular shoveling scene.

Here, \emph{coverage} measures how completely the tool slices through
the material along the target cut plane. It is the fraction of eligible
bins on that plane that show persistent separation on both sides. We use
\emph{simulation coverage} below only for the breadth of scripted material
demonstrations.

\subsection{Tasks}
\label{sec:tasks}

\begin{figure}[t]
\centering
\includegraphics[width=0.95\columnwidth]{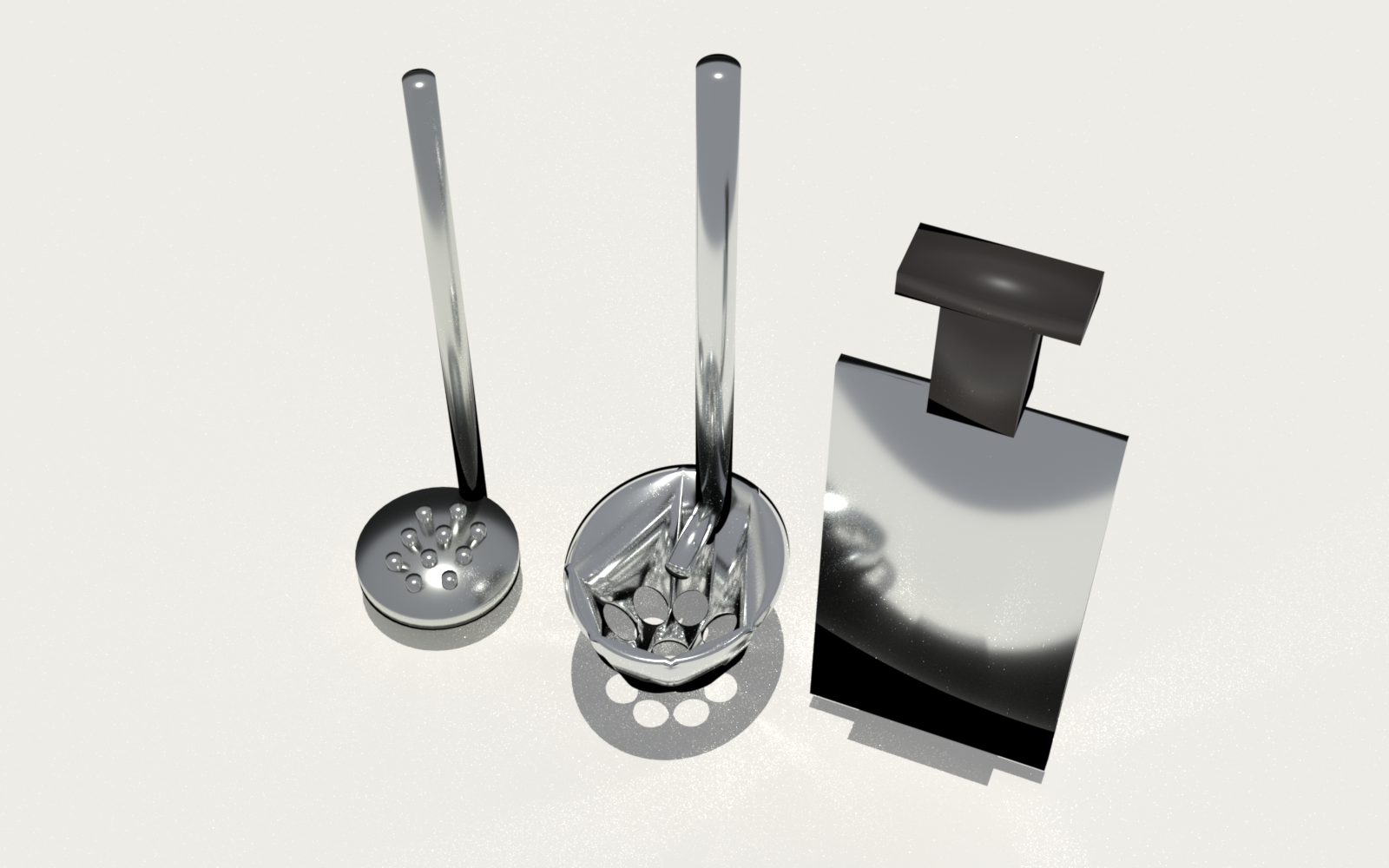}
\caption{The three end-effector tools used in the six
demonstrations of Sec.~\ref{sec:tasks}. Left: a perforated sieve disc
that moves through a granular pile to shovel sand. Centre: an open-bottom
hemispherical scoop that enters the ice-cream tub and lifts out a portion.
Right: a handled knife with a thin planar blade and a wing-style grip,
used for the four cutting demonstrations with jelly, tofu, wax, and
peanut brittle.}
\label{fig:tools}
\end{figure}

We exercise \ours on six demonstrations grouped into three roles:
two RL-trained cutting tasks (\texttt{cut\_tofu},
\texttt{cut\_peanut\_brittle}), two additional cutting demos that
fill in the soft-to-brittle spectrum
(\texttt{cut\_jelly}, \texttt{cut\_wax}), and two non-cutting
demos that exercise the viscoplastic and granular material classes
(\texttt{scoop\_icecream}, \texttt{shovel\_sand}). All six share
the Gym-like observation/action layout of
Sec.~\ref{sec:interface} and the per-particle constitutive layer
of Sec.~\ref{sec:eom}; the three tools (knife, spoon, shovel) used
across the six demos are shown in Figure~\ref{fig:tools}. We narrow
the closed-loop RL evaluation to two representative cutting tasks.
A full \mlsmpm{} substep with the MGPCG damage solve in
$\mathcal{N}(\mathbf{T})$ costs an order of magnitude more wall
time than a comparable rigid-body rollout, so each end-to-end
training run consumes several hours of single-GPU compute even
with the patch optimization of Sec.~\ref{sec:patch}; this scaling is discussed in Sec.~\ref{sec:future}.

\begin{table}[t]
\centering
\caption{Material parameters used by the six demos. All six
demos run the explicit \mlsmpm{}+\ours solver of
Sec.~\ref{sec:three-stage} with $\rho\!=\!1000\,\mathrm{kg/m^3}$
(ice cream: $900$, sand: $1600$). Sand uses a deliberately low
Young's modulus so that the bulk material flows around the tool
in a granular-looking way under the same per-particle damage
layer as the cutting demos.}
\label{tab:materials}
\small
\setlength{\tabcolsep}{4pt}
\begin{tabular}{lccc}
\toprule
Demo & $E$ (Pa) & $\nu$ & $\sigma_F/E$ \\
\midrule
jelly                & $6\!\times\!10^{4}$ & $0.30$ & $0.20$ \\
tofu                 & $2\!\times\!10^{5}$ & $0.40$ & $0.15$ \\
wax                  & $5\!\times\!10^{5}$ & $0.30$ & $0.05$ \\
peanut brittle       & $1\!\times\!10^{6}$ & $0.25$ & $0.015$ \\
ice cream            & $4\!\times\!10^{4}$ & $0.30$ & $0.30$ \\
sand                 & $2\!\times\!10^{3}$ & $0.20$ & $0.10$ \\
\bottomrule
\end{tabular}
\end{table}

\paragraph*{Cutting (RL-trained).}
The two RL-trained cutting tasks (\texttt{cut\_tofu} and
\texttt{cut\_peanut\_brittle}) share the same kinematic setup:
a planar knife held in the FR3 gripper cuts through a slab
on a table. They differ in slab material parameters
(Table~\ref{tab:materials}) and reward weighting.
\texttt{cut\_tofu} uses an extra-firm tofu slab
($E\!=\!200\,\mathrm{kPa}$, $\nu\!=\!0.40$, $\sigma_F/E\!=\!0.15$)
paired with a connectivity-based \emph{cut-quality} reward (clean
band of removed particles with left/right symmetry) and a
\emph{topological split bonus} (weight $60$, fires only when the
alive-particle field has exactly two connected components), with
$200$-step episodes. \texttt{cut\_peanut\_brittle} uses the
candy slab ($E\!=\!1\,\mathrm{MPa}$, $\nu\!=\!0.25$,
$\sigma_F/E\!=\!0.015$): a stiffer, much lower-toughness material
paired with a simpler reward (rate of newly-separated particles
minus proximity and effort penalties), with $250$-step episodes,
so success on \texttt{cut\_peanut\_brittle} is attributable to
the damage solver of Sec.~\ref{sec:three-stage} rather than to
topological-quality shaping. Both tasks share the tip-to-target
proximity term
$\exp(-\lVert\mathbf{x}^{\mathrm{tip}} - \mathbf{x}^{\star}\rVert/0.08)$,
the vertical depth-into-slab term, the blade-axis alignment term
$\cos^2(\hat{\mathbf{n}}_\mathbf{T},-\hat{\mathbf{z}})$, and the
joint-velocity, joint-effort and tool-slip penalties of
Sec.~\ref{sec:interface}. Full reward breakdowns are in
Sec.~\ref{sec:rewards}.

\paragraph*{Cutting (simulation coverage).}
We also render two scripted cutting scenes
with jelly and wax to sample additional material settings
(Table~\ref{tab:materials}). \texttt{cut\_jelly} uses a very soft
gel ($E\!=\!60\,\mathrm{kPa}$, roughly $3\!\times$ softer than
\texttt{cut\_tofu}) with an elevated failure-stress ratio
($\sigma_F/E\!=\!0.20$ vs.\ $0.15$ for \texttt{cut\_tofu}) that
captures the high failure strain of gelatinous solids: the blade
descends through the gel without producing a sharp crack tip and
the two halves wobble apart under their own elasticity, exercising
the regime where damage propagates through elastic shear rather
than brittle fracture. \texttt{cut\_wax} sits between
\texttt{cut\_tofu} and \texttt{cut\_peanut\_brittle} on the same
axis ($E\!=\!500\,\mathrm{kPa}$,
$\sigma_F/E\!=\!0.05$): a stiffer ductile slab whose damage
profile is intermediate between the soft tearing of jelly/tofu and
the clean snap of peanut brittle, confirming that the \cdmpm{}
damage layer interpolates smoothly across stiffness/toughness
combinations rather than only matching the two trained tasks.
Together with the RL-trained pair, these four cutting demos cover
roughly $1.5$ orders of magnitude in stiffness ($60\,\mathrm{kPa}$
to $1\,\mathrm{MPa}$) within the same per-particle damage layer.

\paragraph*{Non-cutting simulation coverage.}
The remaining two demonstrations (\texttt{scoop\_icecream} and
\texttt{shovel\_sand}) exercise the viscoplastic and granular
material classes of Sec.~\ref{sec:constitutive} under scripted
tool motion, both still on the same \mlsmpm{}+\ours backbone
as the cutting demos; only the per-particle material
parameters in Table~\ref{tab:materials} change.
\texttt{scoop\_icecream} (class V) is a semi-fluid lift: a
hemispherical scoop enters an ice-cream tub, carves out a
portion, and lifts it without the chunk fragmenting. The ice
cream is modelled as a very soft elastic substrate
($E\!=\!40\,\mathrm{kPa}$, $\nu\!=\!0.30$,
$\rho\!=\!900\,\mathrm{kg/m^3}$) paired with a high damage
threshold ($\sigma_F/E\!=\!0.30$), so the material flows around
the spoon under sustained shear but the lifted chunk stays
intact below threshold. The same damage solver provides a practical proxy for
Herschel--Bulkley rheology without a separate viscoplastic kernel. \texttt{shovel\_sand}
(class G) drives a perforated sieve disc through a
$30\,\mathrm{cm}\!\times\!30\,\mathrm{cm}\!\times\!10\,\mathrm{cm}$
pile and the grains build a heap in front of the blade. We
model the pile as a heavy ($\rho\!=\!1600\,\mathrm{kg/m^3}$)
near-cohesionless medium: a deliberately low Young's modulus
($E\!=\!2\,\mathrm{kPa}$, $\nu\!=\!0.20$) lets the bulk
particles flow rather than rebound, and a moderate damage
threshold ($\sigma_F/E\!=\!0.10$) lets individual grains detach
from the bulk as they cross the blade rim, so that the
collective behaviour reads as granular under the same
per-particle damage layer used for cutting and scooping.
Together, the four cutting demos and these two non-cutting
scenes confirm that the same Gym-like interface,
observation/action layout, FR3 rigid-body coupling, and
\mlsmpm{}+\ours solver support all four material classes by
adjusting per-particle material parameters alone. Reward
functions and RL training for the four scripted demos
(\texttt{cut\_jelly}, \texttt{cut\_wax}, \texttt{scoop\_icecream},
\texttt{shovel\_sand}) are deferred to
future work (Sec.~\ref{sec:future}).

\subsection{Reward Functions}
\label{sec:rewards}

We express the rewards for the two RL-trained
cutting tasks, \texttt{cut\_tofu} and \texttt{cut\_peanut\_brittle}, as
weighted sums of interpretable shaping terms computed from the \mpm{}
particle field and the rigid-body state. We use the following
notation throughout. $\mathbf{x}^{\mathrm{tip}}\!=\!(x^{\mathrm{tip}},
y^{\mathrm{tip}}, z^{\mathrm{tip}})$ is the tool tip position in
world coordinates and $\hat{\mathbf{n}}_T$ is the tool's body
$\hat{\mathbf{z}}$ axis after orientation. $\mathbf{x}^c\!=\!(c_x,c_y,c_z)$
is the geometric centre of the material block, and
$z_{\mathrm{top}},z_{\mathrm{bot}}$ are the block's top and bottom
$z$-coordinates with height
$h\!=\!z_{\mathrm{top}}\!-\!z_{\mathrm{bot}}$. The FR3
\emph{tool centre point} (TCP) is the standard end-effector
reference frame defined between the gripper fingers, as supplied
by the FR3 URDF;
$z^{\mathrm{TCP}}$ denotes its world $z$, and $z^{\mathrm{knife}}$
denotes the world $z$ of the knife body's centre of mass. When the
gripper closes correctly on the knife handle, the centre of mass
sits $2.25\,\mathrm{cm}$ above the TCP origin
($z^{\mathrm{knife}}\!=\!z^{\mathrm{TCP}}\!+\!0.0225\,\mathrm{m}$),
which is the equilibrium configuration we use as the slip baseline
below. $\dot{\mathbf{q}}\!\in\!\mathbb{R}^{7}$ is the arm
joint-velocity vector. $n^{\mathrm{alive}}_t$ is the number of \mpm{}
particles still active at step $t$,
$\Delta n^{\mathrm{dead}}_t\!=\!n^{\mathrm{alive}}_{t-1}\!-\!n^{\mathrm{alive}}_t$,
and $\Omega^{\mathrm{alive}}_t\!\subset\!\mathbb{R}^{3}$ is the
voxelised support of the alive-particle density at step $t$; we
write $\#\mathrm{cc}(\Omega^{\mathrm{alive}}_t)$ for the number of
$6$-connected components of that voxel set. $\mathbf{a}_t\!\in\!\mathbb{R}^{7}$
is the action emitted by the policy at step $t$ (incremental joint
targets, Sec.~\ref{sec:interface}), and
$\mathbf{a}^{\mathrm{demo}}_t$ is the scripted demonstration
action at the same step (the open-loop sequence of descent, holding, and retraction used for the behavior-cloning warm-start of
Sec.~\ref{sec:implementation}). $\mathbbm{1}[\cdot]$ denotes the
indicator function (1 if the predicate holds, 0 otherwise). Per-task
reward terms below use SI units (metres, radians); all numerical
constants inside indicators and exponentials are in metres unless
stated otherwise, and no global per-step clipping is applied.

\paragraph*{Soft slab reward.}
The \texttt{cut\_tofu} reward decomposes as
\begin{equation}
\begin{aligned}
r^{\mathrm{tofu}}_t \;=\; & 1.5\,r_{\mathrm{reach}} \,+\, 1.0\,r_{\mathrm{depth}} \,+\, 0.5\,r_{\mathrm{vert}} \\
                       & {} \,+\, 15\,r_{\mathrm{cut\text{-}q}} \,+\, 60\,r_{\mathrm{split}} \,+\, 0.5\,r_{\mathrm{track}} \\
                       & {} \,+\, r_{\mathrm{vel}} \,+\, r_{\mathrm{eff}} \,+\, r_{\mathrm{drop}},
\end{aligned}
\label{eq:r-tofu}
\end{equation}
with components
\begin{equation}
\begin{aligned}
r_{\mathrm{reach}} &= \exp\!\bigl(-\lVert\mathbf{x}^{\mathrm{tip}}\!-\!\mathbf{x}^{\star}\rVert/0.08\bigr),\;\;
\mathbf{x}^{\star}\!=\!(c_x,c_y,z_{\mathrm{top}}\!-\!0.02), \\
r_{\mathrm{depth}} &= \min\!\bigl(\max(z_{\mathrm{top}}\!-\!z^{\mathrm{tip}},0),\,h\bigr)/h \in [0,1], \\
r_{\mathrm{vert}} &= \max\!\bigl(0,\;\cos\angle(\hat{\mathbf{n}}_T,\,-\hat{\mathbf{z}})\bigr)^{2}, \\
r_{\mathrm{cut\text{-}q}} &= b_{\mathrm{clear}}\cdot b_{\mathrm{bal}}, \\
r_{\mathrm{split}} &= \mathbbm{1}\!\bigl[\,\#\mathrm{cc}\bigl(\Omega^{\mathrm{alive}}_t\bigr)\!=\!2\bigr], \\
r_{\mathrm{track}} &= \mathbbm{1}\!\bigl[z^{\mathrm{tip}}\!-\!z_{\mathrm{top}}\!<\!0.10\,\mathrm{m}\bigr]\cdot\exp\!\bigl(-2\,\lVert\mathbf{a}_{t-1}\!-\!\mathbf{a}^{\mathrm{demo}}_{t-1}\rVert\bigr), \\
r_{\mathrm{vel}}  &= -0.01\,\sqrt{\lVert\dot{\mathbf{q}}\rVert^{2}\!+\!10^{-8}}, \\
r_{\mathrm{eff}}  &= -2\!\times\!10^{-4}\,\lVert\dot{\mathbf{q}}\rVert^{2}, \\
r_{\mathrm{drop}} &= -2.0\,\max\!\bigl(0,\,(z^{\mathrm{TCP}}\!+\!0.0225)\!-\!z^{\mathrm{knife}}\bigr).
\end{aligned}
\label{eq:r-tofu-comp}
\end{equation}
Here $b_{\mathrm{clear}}\!=\!1\!-\!n^{\mathrm{band}}_t/n^{\mathrm{band}}_0$
measures how much of an axial band of half-width $1.5\,\mathrm{cm}$
centered on the cut plane $y\!=\!c_y$ has been cleared of alive
particles, and $b_{\mathrm{bal}}\!=\!1\!-\!|n_L\!-\!n_R|/n_{\mathrm{far}}$
measures left/right balance of the alive particles outside the
band ($n_L,n_R$ count alive particles with $y\!<\!c_y$ and
$y\!>\!c_y$ respectively). The split bonus $r_{\mathrm{split}}$
counts the connected components of the alive-density volume on a
voxel grid and fires only when there are exactly two, since
"the two halves are balanced" does not imply "the two halves are
topologically separated". An uncut bottom strip can leave a single
component even when the cut quality looks high. The action-tracking
term $r_{\mathrm{track}}$ is gated to the cut zone (tip within
$10\,\mathrm{cm}$ above the block top): without the gate, an early
failure mode is for the policy to mimic the scripted demonstration
while waving the tool in free air, harvesting tracking reward
without ever engaging the material. The two velocity regularizers
act at different scales: $r_{\mathrm{vel}}$ is the main penalty for wrist shaking ($\sim\!\pm 0.1$/step at typical
joint speeds) while $r_{\mathrm{eff}}$ provides a smooth downward
pressure on the action norm without dominating the gradient. The
drop penalty $r_{\mathrm{drop}}$ fires when the knife body's
$z$-coordinate falls below the TCP's $z$ (with the
$2.25\,\mathrm{cm}$ rest offset accounted for). With the
policy-controlled gripper width $a^{\mathrm{grip}}_t$ of
Sec.~\ref{sec:interface}, this term doubles as both a friction-loss
penalty (reactive wrench overcoming the commanded grip) and a
direct deterrent against the policy opening the gripper to drop the
knife: a $5\,\mathrm{cm}$ slip costs $0.1$/step, an order of
magnitude below the cutting reward, so it nudges the policy toward
keeping the tool gripped rather than dominating the learning signal.

\paragraph*{Choice of outer weights and length scales.}
The relative magnitudes of the six shaped weights encode a coarse
curriculum. The dense approach terms
$r_{\mathrm{reach}}, r_{\mathrm{depth}}, r_{\mathrm{vert}}\!\in\![0,1]$
are weighted at $1.5,\,1.0,\,0.5$ respectively: together they
provide the bulk of the early-episode gradient while the tool is
still approaching, and they saturate near $1$ once the blade
reaches the block, at which point $r_{\mathrm{cut\text{-}q}}$ at
weight $15$ becomes the only dense signal that continues to grow
and takes over the gradient. The terminal sparse bonus
$r_{\mathrm{split}}$ is weighted at $60$ so that a single
successful split contributes roughly twice the integrated cut-quality
reward over a $\sim\!200$-step episode, letting topological
success dominate dense near-misses; $r_{\mathrm{track}}$ is held at
$0.5$ because it is an auxiliary descent prior, and a larger
weight would make the policy reproduce the scripted demo even when
material engagement has failed.
The inner length scales follow the workspace geometry. The
$0.08\,\mathrm{m}$ inside $r_{\mathrm{reach}}$ is the radius at
which the exponential reach reward falls to $e^{-1}$, chosen so
the gradient is informative across a $5\!\sim\!10\,\mathrm{cm}$
reach error rather than collapsing to zero. The $0.02\,\mathrm{m}$
offset in $\mathbf{x}^{\star}\!=\!(c_x,c_y,z_{\mathrm{top}}\!-\!0.02)$
places the reach target $2\,\mathrm{cm}$ below the block top, so a
converged policy penetrates the surface rather than hovering
above it. The decay constant $2$ inside the $r_{\mathrm{track}}$
exponential gives half-reward at an action deviation of
$\sim\!0.35\,\mathrm{rad}$/step from the demo, on the same scale
as one \ppo{} update's policy noise. The $1.5\,\mathrm{cm}$
band half-width in $b_{\mathrm{clear}}$ matches the blade
thickness plus a small empirical tolerance for damage spread, so
the band contains exactly the particles that a successful cut
should remove. Finally, the $10^{-8}$ inside the
$r_{\mathrm{vel}}$ square root is a numerical smoothing constant
that keeps the gradient bounded at zero joint velocity and has
no effect on the magnitude of the term elsewhere.

\paragraph*{Hard slab reward.}
The \texttt{cut\_peanut\_brittle} reward is built on the \texttt{cut\_tofu}
shaping above: we keep the joint-effort regularizer (reweighted
$5\!\times$ heavier) and a tip-to-block proximity term, drop the
topological-quality and alignment terms
($r_{\mathrm{cut\text{-}q}}$, $r_{\mathrm{split}}$,
$r_{\mathrm{track}}$, $r_{\mathrm{drop}}$, $r_{\mathrm{vert}}$,
$r_{\mathrm{depth}}$), replace the soft-block reach reward with a
linear distance penalty to the block centre, and add an explicit
per-step kill-count term that rewards the damage solver directly:
\begin{equation}
r^{\mathrm{hard}}_t \;=\; 0.10\,\Delta n^{\mathrm{dead}}_t
\;-\; 0.05\,\lVert\mathbf{x}^{\mathrm{tip}}\!-\!\mathbf{x}^c\rVert
\;-\; 0.001\,\lVert\dot{\mathbf{q}}\rVert^{2}.
\label{eq:r-hard}
\end{equation}
The reduced shaping is intentional: \texttt{cut\_tofu}'s
$r_{\mathrm{cut\text{-}q}}$ and $r_{\mathrm{split}}$ can in
principle be earned by any policy that cleanly displaces the band
of particles between the two halves, regardless of whether the
displacement comes from \cdmpm{} damage propagation or from raw
particle-deactivation artefacts. Stripping those terms on the
stiffer slab and rewarding particle deactivation directly through
$\Delta n^{\mathrm{dead}}_t$ attributes success unambiguously to
the damage solver of Sec.~\ref{sec:three-stage}, which is what we
ablate in Sec.~\ref{sec:baselines}. The three coefficients of
$r^{\mathrm{hard}}_t$ are calibrated so that the kill-count term
dominates while the regularizers remain secondary: at
$\Delta n^{\mathrm{dead}}_t\!\sim\!10$ particles/step typical of a
working cut, $0.10\,\Delta n^{\mathrm{dead}}_t$ contributes
$\sim\!+1$/step, against $\sim\!-0.025$/step from the
$0.05\lVert\mathbf{x}^{\mathrm{tip}}\!-\!\mathbf{x}^c\rVert$
centre-distance term at a typical $0.5\,\mathrm{m}$ workspace
distance and $\sim\!-0.01$/step from the joint-effort term. The
joint-effort coefficient $0.001$ is $5\!\times$ heavier than
\texttt{cut\_tofu}'s $r_{\mathrm{eff}}\!=\!-2\!\times\!10^{-4}$ to
account for the larger joint torques required to drive the
stiffer slab, which would otherwise let brute-force motion
substitute for controlled cutting.

\subsection{Implementation and Hyperparameters}
\label{sec:implementation}

\paragraph*{Simulator and robot stack.}
All physics runs on NVIDIA \textsc{Warp}~\cite{warp2022} as a
unified GPU backend. The \cdmpm{} solver of
Sec.~\ref{sec:three-stage} discretizes the equations of motion
with explicit \mlsmpm{}~\cite{HuFangGeEtAl18}: mass and momentum are
transferred to a fixed background grid with $\sim\!16\mathrm{K}$
active cells via APIC, the grid update is a single Warp pass per
substep (no global linear solve), and the per-particle damage
update of Sec.~\ref{sec:three-stage} is a local closed-form
Cardano eigenvalue of the symmetric stress tensor. The articulated
FR3 arm and the free-rigid tool are stepped by a Warp-resident
MuJoCo solver (\texttt{mujoco\_warp}) that shares the device and
CUDA stream with \mpm{}, so the two-way coupling kernels of
Sec.~\ref{sec:coupling} (\texttt{compute\_body\_forces},
\texttt{subtract\_body\_force}) exchange impulses on-device without
any host round-trip. Control runs at $50\,\mathrm{Hz}$, with
per-control \mpm{} and rigid substeps tuned per task to satisfy the
Courant--Friedrichs--Lewy (CFL) stability condition:
\texttt{cut\_tofu} uses voxel size $0.04\,\mathrm{m}$ with
$1\!+\!1$ substeps ($\sim\!19\mathrm{K}$ particles per env),
\texttt{cut\_peanut\_brittle} voxel $0.018\,\mathrm{m}$ with $3\!+\!3$
substeps, \texttt{scoop\_icecream} voxel $0.018\,\mathrm{m}$ with
$2\!+\!2$ substeps, and \texttt{shovel\_sand} voxel
$0.020\,\mathrm{m}$ with $2\!+\!2$ substeps. The tool is held by
contact friction in the FR3 fingers (gripper PD gains
$k_p\!=\!5\!\times\!10^{4}$, $k_d\!=\!700$, friction $\mu\!=\!2.0$)
rather than welded to the end-effector, which is what makes the
gripper-slip penalty in Sec.~\ref{sec:interface} a meaningful
signal. The production damage solver uses outer PCG with a full
geometric V-cycle preconditioner, a relative-residual tolerance of
$10^{-5}$, and a cap of 20 outer iterations. Each V-cycle applies three
weighted-Jacobi pre- and post-smoothing sweeps and 24 coarse-level
iterations; restriction and prolongation remain on the GPU.

\paragraph*{Rendering.}
Both rendering paths share the same geometry pipeline: alive \mpm{}
particles are voxelized into a density field, Gaussian-smoothed,
and converted into a triangle mesh by marching cubes. The
lightweight path rasterizes the mesh through Newton's Warp-resident
viewport for training-time inspection, while the offline path
imports the mesh into Mitsuba~3~\cite{mitsuba3} for Monte~Carlo
path-traced figures with physically based area lighting and a
smooth-dielectric model for the brittle-jelly material.

\paragraph*{Two-stage training pipeline with MPM-derived force jittering.}
The full \mlsmpm{} environment limits training throughput: a single env runs
at $\sim\!12\,\mathrm{fps}$ for raw simulation and $\sim\!5\text{--}6\,\mathrm{fps}$
with \ppo{} updates included. Training $2\!\times\!10^{5}$
on-policy steps at this throughput requires substantial wall-clock time. We
exploit the observation that the gripper-and-arm sub-problem (track
a target with the tool firmly held, reject an unknown end-effector
load) is a pure rigid-body control problem and can be pre-trained
on a much cheaper proxy environment. Concretely, we use a
two-stage pipeline.

\emph{Stage~1: rigid-body pre-training with \mpm{}-calibrated wrench
jitter.}
We first run a small set of scripted \mlsmpm{} rollouts of the target
task and collect, via \texttt{mpm.collect\_collider\_impulses}, the
per-step net force $\mathbf{F}^{\mathrm{mpm}}_t$ and torque
$\boldsymbol{\tau}^{\mathrm{mpm}}_t$ that the cutting material
applies to the tool. From these we estimate the empirical 95th
percentile of the force and torque magnitudes, written
$p_{95}^{F}$ and $p_{95}^{\tau}$: that is, in $95\%$ of simulation
steps $|\mathbf{F}^{\mathrm{mpm}}_t|\!\le\!p_{95}^{F}$, and
analogously for the torque. We use the 95th percentile rather than
the maximum because the maximum is dominated by transient
impact and fracture spikes that misrepresent the typical contact
load, and rather than the mean because the mean is pulled down by
the many steps in which the tool is not engaged with the material;
on the cutting tasks the estimates are $p_{95}^{F}\!\approx\!0.5\,\mathrm{N}$
and $p_{95}^{\tau}\!\approx\!5\,\mathrm{N\!\cdot\!m}$. We then train
on a pure rigid-body environment built on \texttt{mujoco\_warp},
with no \mpm{} at all, in which at every control step we sample
each Cartesian component of the applied force and torque
independently and uniformly within these calibrated bounds:
\begin{equation}
\mathbf{F}^{\mathrm{jit}}_t\!\sim\!\mathcal{U}\!\bigl([-p_{95}^{F},+p_{95}^{F}]^{3}\bigr),\;\;
\boldsymbol{\tau}^{\mathrm{jit}}_t\!\sim\!\mathcal{U}\!\bigl([-p_{95}^{\tau},+p_{95}^{\tau}]^{3}\bigr),
\label{eq:jitter}
\end{equation}
and apply the resulting wrench to the tool body for that step.
This injects a disturbance with the same characteristic magnitude
that \mpm{} contact will eventually deliver, without running the \mpm{} simulation. Because the entire stage avoids \mpm{}, we batch
$\sim\!8192$ FR3+tool worlds inside a single \texttt{mujoco\_warp}
\texttt{nworld} model via \texttt{newton.ModelBuilder.replicate()}
and train on \emph{one} GPU; observation and action layouts match
Eqs.~\eqref{eq:obs}--\eqref{eq:action} exactly so the resulting
policy is checkpoint-compatible with Stage~2.

\emph{Stage~2: \ours fine-tuning.}
We load the Stage~1 checkpoint with the \texttt{--init-checkpoint}
flag and continue \ppo{} inside the full \ours{} environment of
Sec.~\ref{sec:method}. This environment combines the
tool-centered MGPCG damage solve with the \mlsmpm{} momentum
substrate. We use a single-env, single-process Warp rollout on the same GPU. This is where the policy specializes
from "track a target while rejecting an unknown end-effector load"
to "engage the material, drive damage along the desired plane,
release on completion." Stage~2 uses
$4\!\times\!10^{4}\text{--}10^{5}$ \ours{} steps to fine-tune the Stage~1 policy.

\paragraph*{Networks.}
\ppo{}'s actor-critic is a two-hidden-layer MLP of width $256$ with
$\tanh$ activations and orthogonal initialization (output-layer
gain $0.01$ on the policy head, $1.0$ on the value head). The
policy is a diagonal Gaussian with a learnable, state-independent
$\log\sigma$ initialized at $-3$ so the initial policy stays close
to zero. Observations are normalized online by a running
mean/variance estimate updated at every rollout step; without this
normalization the joint angles ($\sim\!3\,\mathrm{rad}$) dominate
target-offset components ($\sim\!5\,\mathrm{cm}$) and the policy
fails to learn the millimetre-scale corrections demanded by the
cutting tasks.

\paragraph*{PPO.}
We use the clip variant of \ppo{}~\cite{SchulmanWolskiDhariwalEtAl17}
with rollout length $2048$ steps per env, $10$ update epochs,
minibatch size $256$, learning rate $3\!\times\!10^{-4}$ (Adam,
default betas), clip range $\epsilon\!=\!0.2$, value coefficient
$0.5$, no entropy bonus, $\mathrm{GAE}(\gamma\!=\!0.99,\,
\lambda\!=\!0.95)$, gradient-norm clipping at $0.5$, and a
target-KL early stop at $0.02$ that aborts the remaining update
epochs if the empirical KL of the last epoch exceeds threshold.
Advantages are normalized per update across the
$T\!\times\!N$ rollout; value-loss clipping is not used. Single
batches whose gradients contain non-finite entries (occasional on
the cutting tasks when a fracture event injects a spike into the
reward) are detected and silently skipped to avoid invalid optimizer updates.

\paragraph*{PPO variants.}
On the cutting tasks (\texttt{cut\_tofu},
\texttt{cut\_peanut\_brittle}) a single-environment \ppo{} actor
trained from a random initialization collapses to a do-nothing
return in $[-27, +27]$: random Gaussian on-policy exploration
almost never produces a clean sequence of descent, engagement, and cutting, and
the advantage signal is overwhelmed by noise. We address this
through two variants. \emph{(i)~Vectorized \ppo{}} rolls out $64$
parallel environments and aggregates the gradient across them,
stabilizing training to a mean return of $+328$ on
\texttt{cut\_tofu} with $\sim\!2600$ particles removed per episode;
this is the strongest cutting policy we report. \emph{(ii)~\ppo{}
with BC warm-start} initializes a single-environment \ppo{} actor
from the behavior-cloning prior described next, yielding a
deterministic return of $\sim\!+278$ that subsequent \ppo{}
refinement preserves.

\paragraph*{Behavior cloning (BC) and critic warm-start.}
Stage~1 itself begins with a behavior-cloning pass that initializes
the policy from a hand-scripted action sequence. The scripted
trajectory is a deterministic sequence of descent, holding, and retraction for the
cutting tasks and analogous hand-coded primitives for shoveling
and scooping; on \texttt{cut\_tofu} it reliably produces an
episode return of $\sim\!+325$, while a freshly-initialized
random policy plateaus at $+100\!\sim\!+150$ because random
Gaussian on-policy actions almost never form a clean
sequence of descent, engagement, and cutting and the critic never sees an example
of success. We use the demo for two purposes. First, an
\emph{actor BC pass} (20 epochs at learning rate $10^{-3}$) fits
the diagonal-Gaussian policy by minimizing the negative
log-likelihood
$-\sum_t\log\pi_\theta(\mathbf{a}^{\mathrm{demo}}_t\mid\mathbf{o}_t)$
on a dataset of $\sim\!4{,}000$--$64{,}000$ demonstration tuples
collected by replaying the scripted action across $64$ parallel
rigid-body envs; the \mpm{}-calibrated wrench jitter of Stage~1
gives every env a different perturbation history, providing the
observation-side diversity that lets BC generalize beyond the
exact demo trajectory rather than collapsing onto it. After BC,
the policy's per-dimension $\log\sigma$ is floored at $-2$
($\sigma\!\approx\!0.13$): without this floor NLL drives
$\log\sigma$ to very negative values on near-deterministic action dimensions,
and the first \ppo{} update's empirical KL diverges. Second, an
optional \emph{critic BC} (``BC-critic'') fits the value head on
the same demonstration data to the Monte-Carlo discounted return,
$V_\phi(\mathbf{o}_t)\!\approx\!\mathbb{E}\!\left[\textstyle\sum_{k\ge 0}\gamma^k r_{t+k}\,\big|\,\mathbf{o}_t\right]$,
with $\gamma$ matched to the \ppo{} discount. Without this prefit
the critic is random and the first \ppo{} iteration produces
advantages $A\!=\!R-V$ of the same order as the raw return, which
once again causes a sharp increase in KL divergence and destabilizes the BC-initialized actor;
with the prefit the iter-1 advantage stays bounded and \ppo{}
refines the BC policy stably. The combined procedure (BC actor
$+$ BC critic $+$ $\log\sigma$ floor) is what produces the
``\ppo{} with warm-start'' rows in the comparison below.

\paragraph*{Compute budget.}
All experiments in this paper were run on a single NVIDIA
RTX~4090 GPU; both stages share the same device. Stage~1, despite hosting
$8192$ rigid-body envs in a single \texttt{mujoco\_warp}
\texttt{nworld}, fits on one GPU because the model
contains no \mpm{}. Stage~2 is the throughput bottleneck: a single
\mlsmpm{} env runs at $\sim\!12\,\mathrm{fps}$ for raw simulation
(scripted action, no policy) and $\sim\!5\text{--}6\,\mathrm{fps}$
once policy inference, GAE, and the \ppo{} update are layered on; the
Stage~2 budget is $4\!\times\!10^{4}\text{--}10^{5}$ \mpm{} steps per
$(\text{task},\text{seed})$. We train $3$ seeds per configuration
and report mean$\,\pm\,$one standard deviation; checkpoints are
written every $5$ outer iterations and the best running mean
return is preserved.

\subsection{Baselines and Ablations}
\label{sec:baselines}

We evaluate four aspects: solver speed at matched residual, momentum
conservation in a free collision, policy sensitivity to damage modeling, and
crack quality across tools and materials. Unless noted otherwise, timings
use one NVIDIA RTX~4090.

\subsubsection{Localized MGPCG Efficiency}
\label{sec:solver-eval}

We separate localization from multigrid acceleration by solving the
same 3D implicit phase-field system in four ways: full-grid Jacobi-PCG,
localized Jacobi-PCG, localized PCG with a symmetric two-level
preconditioner, and localized PCG with a full V-cycle (our MGPCG solver).
All four use the same right-hand side, initial guess, operator, and
relative-residual tolerance $10^{-5}$. Five timed CUDA repeats include all
kernels and patch transfers. Table~\ref{tab:solver-ablation} reports the
times and iteration counts, separating localization from preconditioning.

\begin{table*}[t]
\centering
\caption{Matched-residual 3D damage-solver ablation. Time is median
milliseconds per solve.}
\label{tab:solver-ablation}
\scriptsize
\begin{tabular}{crrrrrrrr}
\toprule
Grid & \multicolumn{2}{c}{Full PCG} & \multicolumn{2}{c}{Local PCG} &
\multicolumn{2}{c}{Two-level PCG} & \multicolumn{2}{c}{Local MGPCG (Ours)}\\
 & ms & iter. & ms & iter. & ms & iter. & ms & iter.\\
\midrule
$192^3$ & 139.38 & 149 & 40.15 & 144 & 54.45 & 11 & 35.80 & 4\\
$256^3$ & 694.36 & 199 & 53.90 & 195 & 73.96 & 15 & 35.98 & 4\\
$320^3$ & 1741.15 & 249 & 115.57 & 246 & 105.61 & 19 & 38.60 & 4\\
\bottomrule
\end{tabular}
\end{table*}

Localization alone yields $3.47\times$, $12.88\times$, and
$15.07\times$ speedups at $192^3$, $256^3$, and $320^3$; full multigrid
preconditioning raises these to $3.89\times$, $19.30\times$, and
$45.11\times$. At the default patch the localized field differs from the
full reference by $0.65\%$. Increasing patch half-width from
$3.75\ell_c$ to $7.5\ell_c$ reduces error from $1.61\%$ to $0.074\%$,
while solver time increases from 35.0 to 48.4 ms. Thus localization is
accurate when its boundary is separated from the active damage band, not
necessarily exact in every scene.

In the production Warp solver, the same 225-frame, 806k-particle
tofu trajectory reaches median residual $2.83\!\times\!10^{-6}$ (maximum
$9.87\!\times\!10^{-6}$) in 8 MGPCG iterations, costing 319.56 ms for
MPM and 330.44 ms total per control frame. The previous fixed three-cycle
path costs 272.74 ms but stops at median residual
$1.06\!\times\!10^{-2}$; it is faster only because it is under-solved.
The final damage fields differ by $4.56\%$ in relative $L_2$.

\subsubsection{Two-Way Coupling Conservation}

For conservation, we remove gravity, controls, and the floor. A
translating free \mpm{} block collides with a stationary 2-kg free rigid box
using the production SDF contact and impulse-accumulation path. We use a
free body rather than an FR3-held tool because the robot's fixed base, joint
constraints, and actuators exchange external impulse with the world. The
free-body setup therefore isolates the equal-and-opposite exchange in the
coupling kernels. The block travels at $1.5\,\mathrm{m/s}$ along $+x$.
In the \emph{centered} case, its impulse line passes through the box center
of mass, up to a $10^{-4}\,\mathrm m$ offset that avoids an ambiguous
symmetric SDF face. This mainly tests linear-momentum exchange. In the
\emph{eccentric} case, the block center is displaced by $3.5\,\mathrm{cm}$
in $y$, which adds a moment arm and tests angular as well as linear
momentum. Twelve runs combine these contacts with grid spacings
$0.012/0.010/0.008$ m and timesteps $2\!\times\!10^{-4}/10^{-4}$ s.
The audit uses the CPU backend of the same kernel and has no external
impulse.

At each step we compute total linear momentum $\mathbf P$ and
angular momentum $\mathbf L$ about the world origin. $\mathbf L$ includes
the \mpm{}/APIC orbital and affine terms, plus rigid-body orbital and spin
terms. We report
$e_P=\lVert\mathbf P(t)-\mathbf P(0)\rVert/P_s$ and
$e_L=\lVert\mathbf L(t)-\mathbf L(0)\rVert/L_s$, with
$P_s=\max(\lVert\mathbf P(0)\rVert,\epsilon)$ and
$L_s=\max(\lVert\mathbf L(0)\rVert,
\lVert\mathbf P(0)\rVert\max(\lVert\mathbf x_T(0)\rVert,0.1\,\mathrm m),
\epsilon)$. The stepwise columns use the corresponding residuals
$\Delta\mathbf p_{\mathrm{MPM}}+\mathbf J_T$ and
$\Delta\mathbf L_{\mathrm{MPM}}+\mathbf x_T\!\times\!\mathbf J_T+\mathbf K_T$,
with the same normalization.

\begin{table}[H]
\centering
\caption{Maximum normalized conservation error. Step $P/L$ measures
equal-and-opposite contact impulse.}
\label{tab:conservation}
\scriptsize
\begin{tabular}{lrrrr}
\toprule
Contact & $e_P$ & $e_L$ & Step $P$ & Step $L$\\
\midrule
Centered & $1.06$e--6 & $8.35$e--7 & $3.36$e--7 & $2.67$e--7\\
Eccentric & $1.53$e--6 & $1.33$e--6 & $6.50$e--7 & $5.69$e--7\\
\bottomrule
\end{tabular}
\end{table}

Table~\ref{tab:conservation} reports the maximum error over all grid
spacings, timesteps, and simulated steps. Both the global conservation
errors and per-step action--reaction residuals remain at approximately
$10^{-6}$ or below. The eccentric case is the stricter angular-momentum
test because of its nonzero lever arm; its maximum $e_L$ is
$1.33\!\times\!10^{-6}$. These results validate equal-and-opposite linear
and angular impulse exchange to numerical precision. Angular impulse is
accumulated about the rigid body's center of mass; an articulated-body
bridge must shift the moment if its collider reference differs from it.

\subsubsection{Does Damage Fidelity Improve the Policy?}

We construct paired compact Gym ablations. They share the same
observation definitions, actions, rewards, initial particles, robot and
\mpm{} configurations, PPO network, BC warm start, 10,240-step budget, and
three seeds. Their observations use the same robot-state-plus-task-summary
interface as the preceding tasks, with scalars for cut completion,
tip-to-center distance, tip height, cut-plane error, and contact force.
The internal damage model is the only change. The \emph{full} model uses
stress history, localized phase-field MGPCG, damage-dependent stress
degradation, and thresholded CPIC separation. The \emph{simplified} model
assigns persistent CPIC sides directly from the blade sweep. Both models use
the same randomly positioned particle region intersecting the target plane,
but only the full model represents its increased fracture resistance; the
simplified model is material-agnostic. The inclusion location is hidden from
the policy, although contact-force feedback remains visible. For this
ablation, we divide the nominal $y=0$ cut plane into $1.5$-cm $x$--$z$ bins
and retain bins that initially contain particles on both sides. We count a
bin as covered when persistent CPIC labels appear on both sides. Success
requires at least $80\%$ coverage and cut-plane error at most 2.5 cm.

\begin{table}[t]
\centering
\caption{Three-seed policy cross-evaluation (mean$\pm$SD).}
\label{tab:policy-ablation}
\scriptsize
\begin{tabular}{llrr}
\toprule
Train & Test & Success & Coverage\\
\midrule
Full & Full & 3/3 & $84.85\!\pm\!0.66\%$\\
Full & Simplified & 3/3 & $100.0\!\pm\!0.0\%$\\
Simplified & Full & 2/3 & $76.89\!\pm\!13.46\%$\\
Simplified & Simplified & 3/3 & $100.0\!\pm\!0.0\%$\\
\bottomrule
\end{tabular}
\end{table}

Both policies solve the simplified test, but the full-trained policy
has higher success and coverage under heterogeneous full physics.
This pilot study provides a controlled comparison across three random
seeds under the same training budget.
The compact ablation
welds the tool to isolate separation fidelity and therefore does not
measure gripper slip.

\subsubsection{Tool--Material Crack Quality}

To test fracture fidelity beyond one example, we combine three
controlled 3D tool-shaped driving fields (thin straight, thick straight,
and wedge) with soft/tough, medium, and hard/brittle materials. For each
pair, full-grid PCG and localized MGPCG solve the same phase-field system to
$10^{-5}$. This is a solver validation, not nine end-to-end robot rollouts.
Table~\ref{tab:crack-matrix} reports damage-field, fracture-energy, and
crack-surface discrepancies.

\begin{table}[t]
\centering
\caption{Crack-quality matrix. $L_2$ and $E_f$ are relative damage-
field and fracture-energy errors; $H$ is surface Hausdorff distance.}
\label{tab:crack-matrix}
\scriptsize
\begin{tabular}{llrrr}
\toprule
Material & Tool & $L_2$ (\%) & $E_f$ (\%) & $H$ (mm)\\
\midrule
Soft/tough & Thin & 0.132 & 1.196 & 0.0\\
 & Thick & 0.235 & 1.285 & 2.5\\
 & Wedge & 0.193 & 0.877 & 2.5\\
Medium & Thin & 0.114 & 0.664 & 2.5\\
 & Thick & 0.200 & 0.722 & 2.5\\
 & Wedge & 0.154 & 0.435 & 2.5\\
Hard/brittle & Thin & 0.088 & 0.302 & 2.5\\
 & Thick & 0.153 & 0.334 & 2.5\\
 & Wedge & 0.108 & 0.171 & 2.5\\
\bottomrule
\end{tabular}
\end{table}

Across all combinations, damage-field error is
$0.088\%$--$0.235\%$, median fracture-energy error is $0.66\%$ and the
maximum is $1.29\%$, and maximum Hausdorff distance is one 2.5-mm cell.
Soft/tough material under the thin field produces no thresholded surface
in either solver and is recorded as a consistent no-crack result. Maximum
separated-volume discrepancy is $2.1\!\times\!10^{-5}$ of the domain.
Thus, across this tool--material matrix, restricting the solve to the
tool-centered patch preserves the full-grid crack field, fracture energy,
and extracted crack surface to within the reported discretization errors.

\subsubsection{Two-Dimensional CD-MPM Comparison}

\begin{figure*}[t]
\centering
\includegraphics[width=0.88\textwidth]{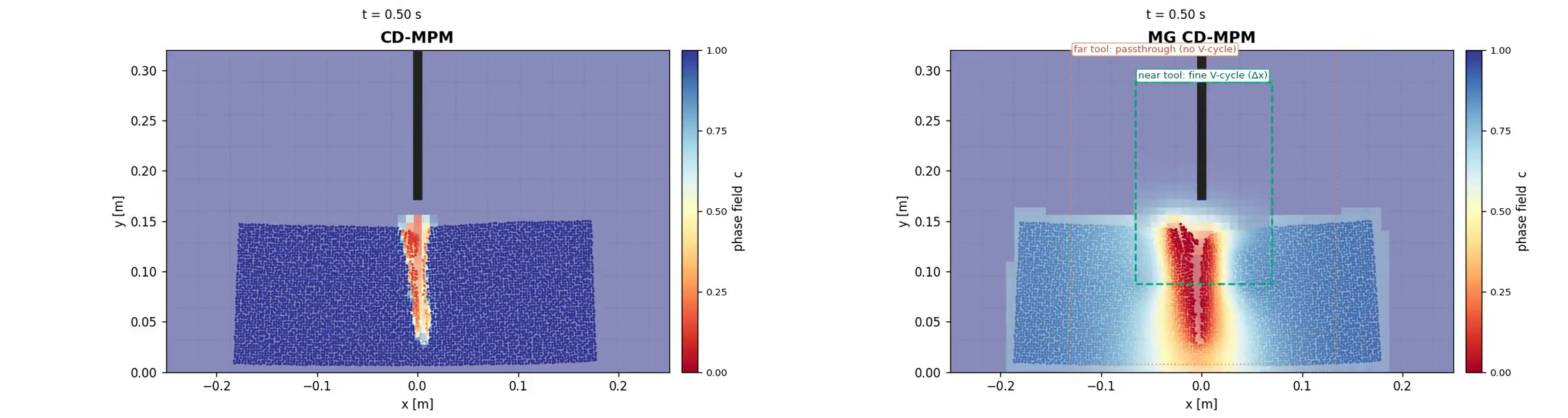}
\caption{Phase field $c$ at $t=0.50\,\mathrm{s}$ in a 2D cutting
example on a $63\!\times\!41$ uniform background grid. Left: the
\cdmpm{} reference. Right: the localized fixed-V-cycle configuration; the
green dashed box is the $16\!\times\!24$ active solve region. The outer
orange dotted box is shown only as far-field context and is not a second
solve level.}
\label{fig:cdmpm-baseline-revised}
\end{figure*}

The localized phase-field construction also applies in two
dimensions. Figure~\ref{fig:cdmpm-baseline-revised} shows a 2D cutting
example at $t=0.50\,\mathrm{s}$ on a $63\!\times\!41$ uniform grid. The
left panel is the \cdmpm{} reference. The right panel applies the phase-field
V-cycle only in the tool-centered $16\!\times\!24$ region and keeps the
field outside it. This produces a spatially regularized damage band. Because
the example uses a prescribed number of V-cycles, it is for qualitative
visualization rather than matched-accuracy timing or speedup claims; the
controlled 3D comparison appears in Table~\ref{tab:solver-ablation}.

\paragraph*{Simulator baselines.}
The \emph{Full-PCG} baseline implements the original \cdmpm{}
phase-field system on the complete uniform grid with Jacobi-preconditioned
CG. It shares our particle state, constitutive update, transfers, contact,
right-hand side, and residual tolerance; only damage localization and the
preconditioner differ. This is the reference used in
Table~\ref{tab:solver-ablation} and the 3$\times$3 crack matrix.

\begin{figure*}[t]
\centering
\begin{minipage}{0.245\textwidth}\centering
\includegraphics[width=\textwidth]{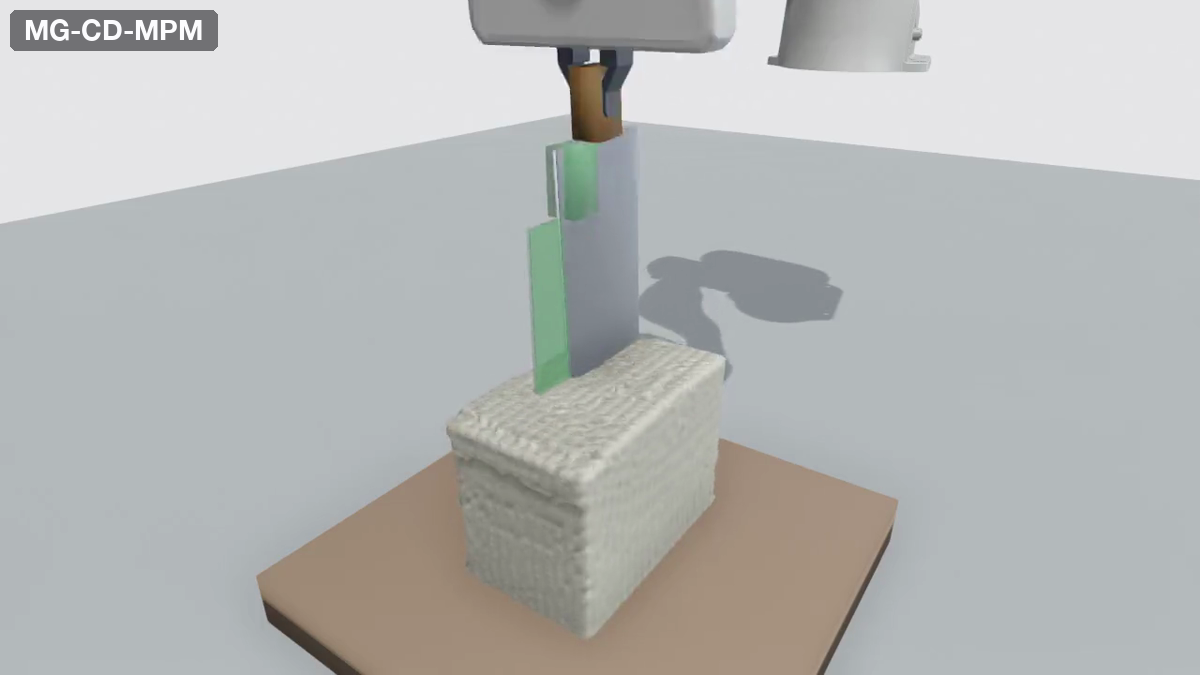}
\end{minipage}\hfill
\begin{minipage}{0.245\textwidth}\centering
\includegraphics[width=\textwidth]{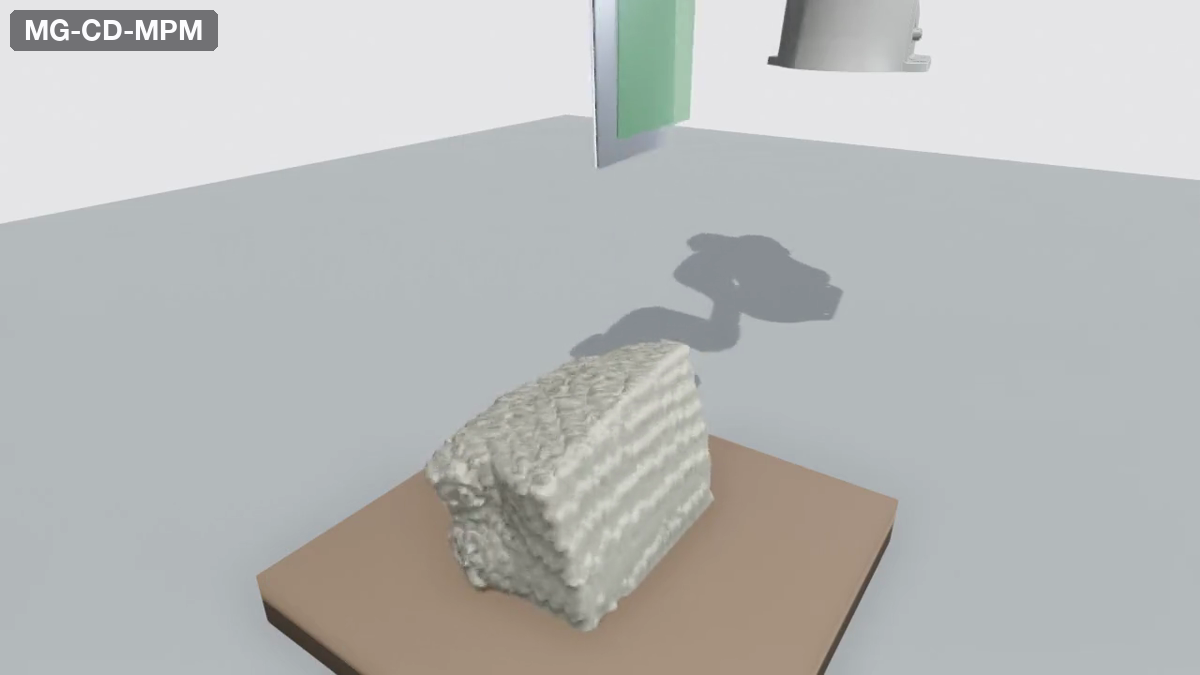}
\end{minipage}\hfill
\begin{minipage}{0.245\textwidth}\centering
\includegraphics[width=\textwidth]{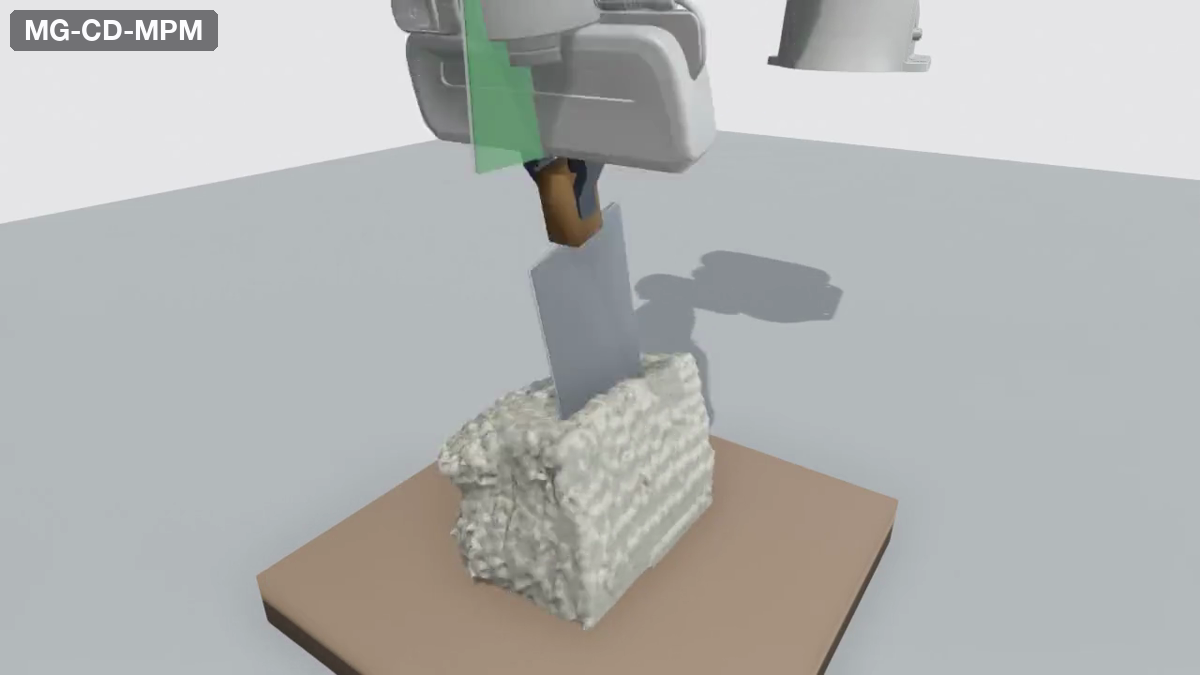}
\end{minipage}\hfill
\begin{minipage}{0.245\textwidth}\centering
\includegraphics[width=\textwidth]{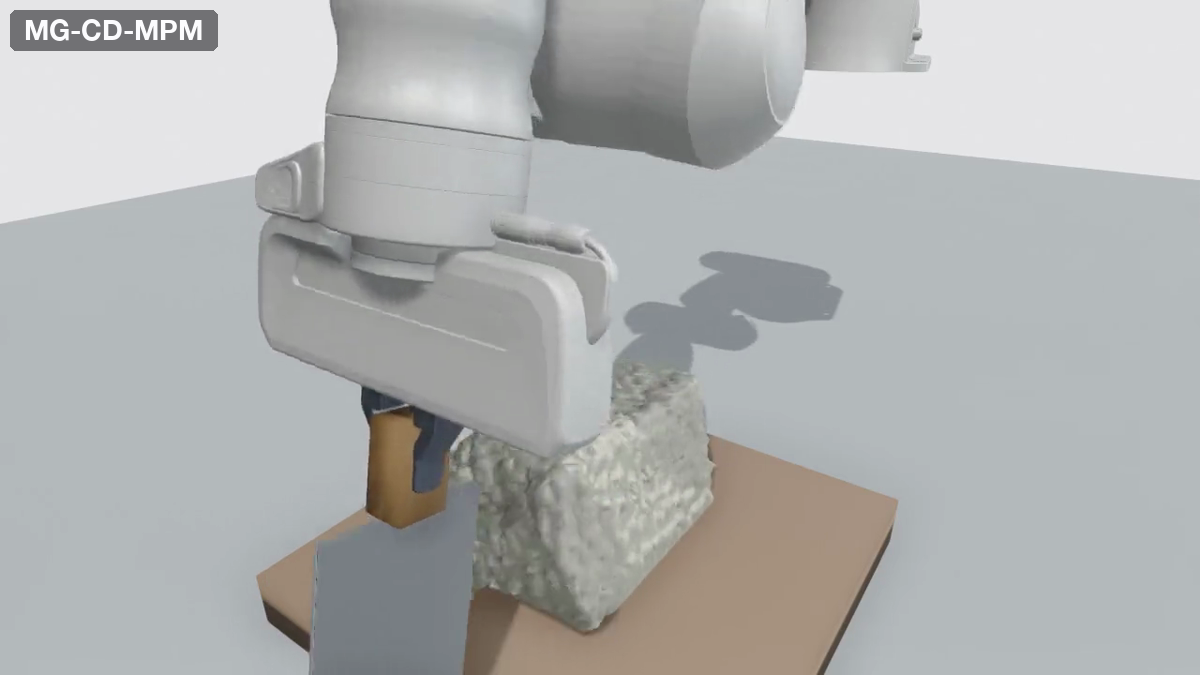}
\end{minipage}\\[2pt]
\begin{minipage}{0.245\textwidth}\centering
\includegraphics[width=\textwidth]{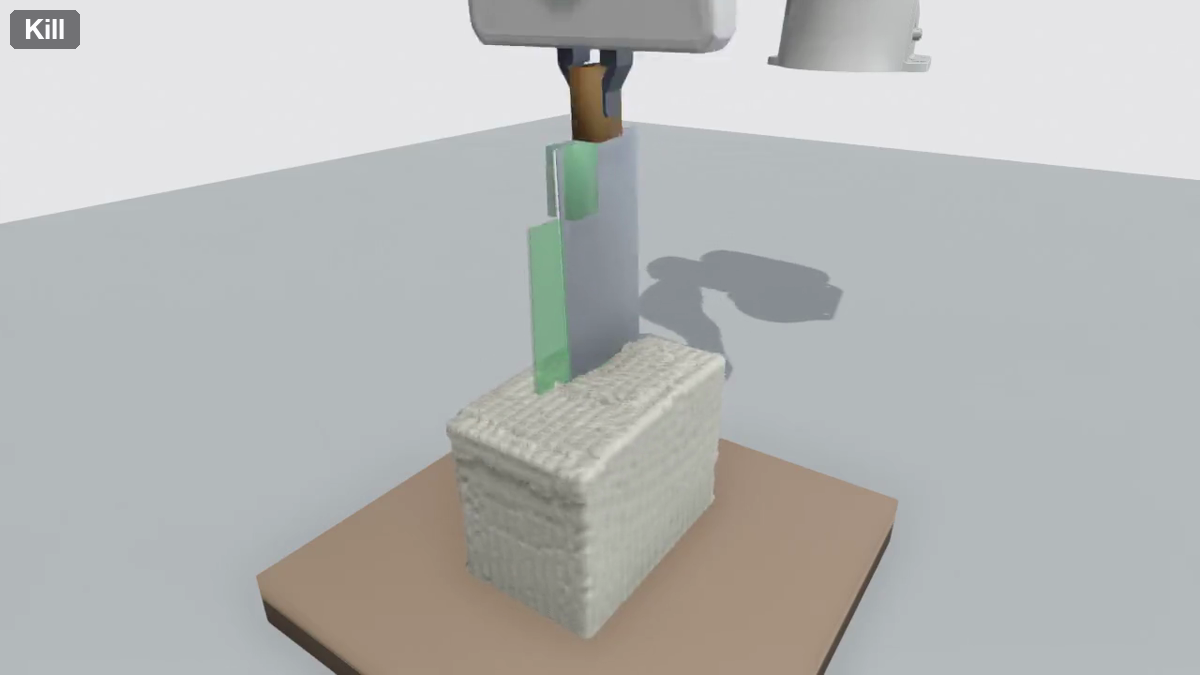}\\[-2pt]
{\scriptsize $t=0.5\,\mathrm{s}$}
\end{minipage}\hfill
\begin{minipage}{0.245\textwidth}\centering
\includegraphics[width=\textwidth]{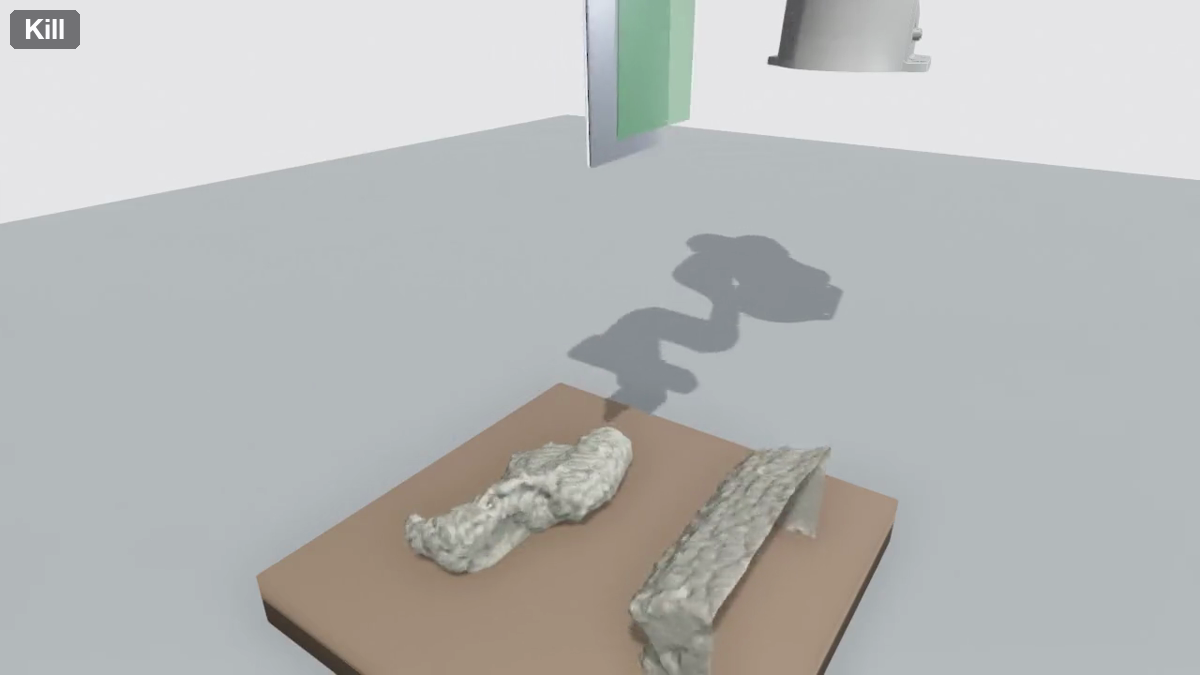}\\[-2pt]
{\scriptsize $t=1.5\,\mathrm{s}$}
\end{minipage}\hfill
\begin{minipage}{0.245\textwidth}\centering
\includegraphics[width=\textwidth]{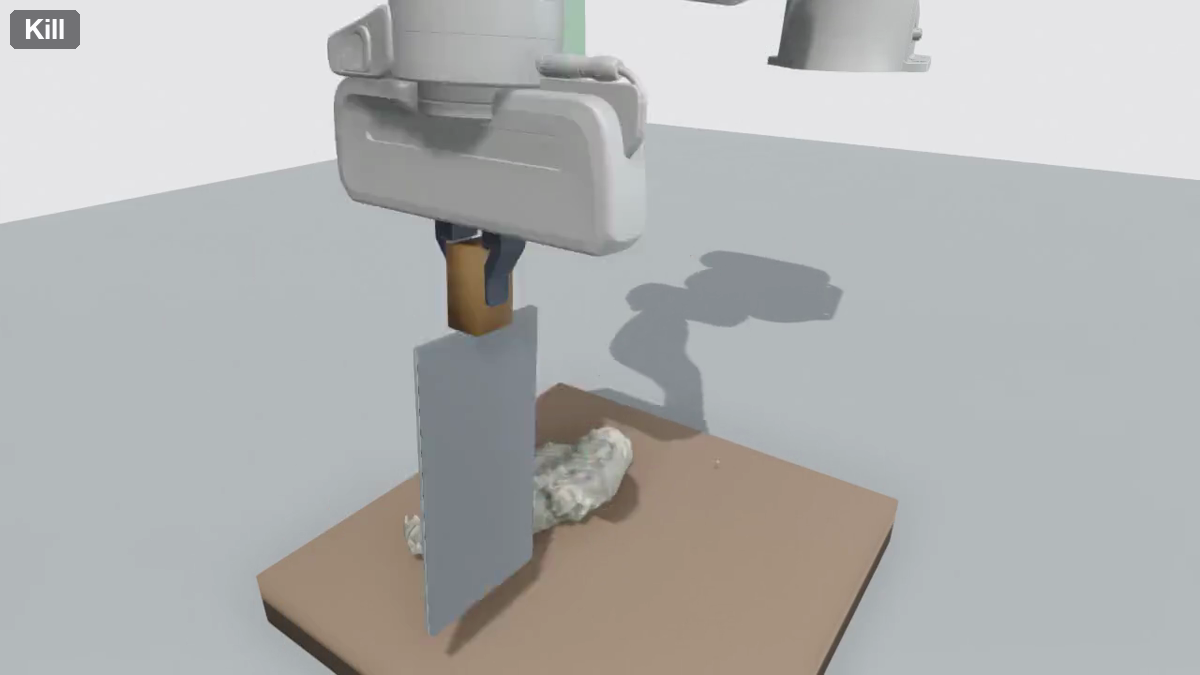}\\[-2pt]
{\scriptsize $t=2.5\,\mathrm{s}$}
\end{minipage}\hfill
\begin{minipage}{0.245\textwidth}\centering
\includegraphics[width=\textwidth]{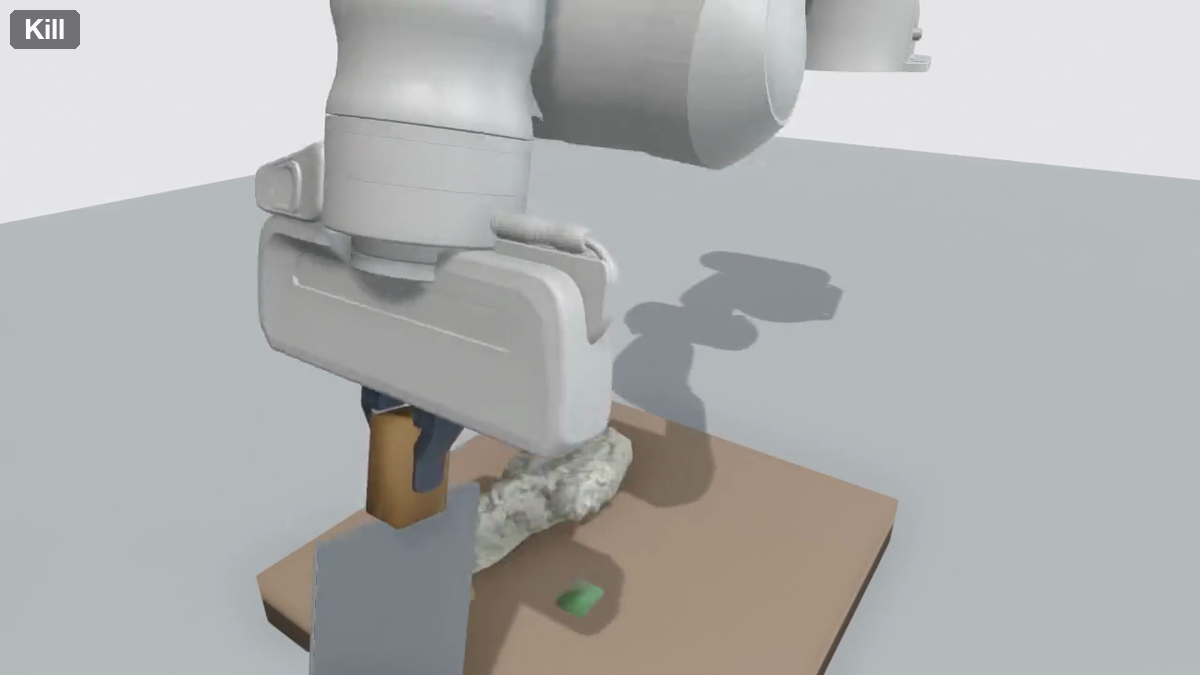}\\[-2pt]
{\scriptsize $t=3.5\,\mathrm{s}$}
\end{minipage}
\caption{Replay on a tofu block of the stable RL trajectory learned
with our phase-field environment. Top: coherent phase-field separation.
Bottom: the Kill mask applied to the same replay; a direct Kill-based RL
rollout diverges at blade contact.}
\label{fig:kill-revised}
\end{figure*}

\emph{Kill} is the simplest legacy rule: particles swept by the blade are
removed from subsequent P2G/G2P transfers. It deletes mass and stored
elastic energy, produces sparse grid mass and impulse spikes, and is not a
conservative or stable RL environment. Standalone Kill diverges at
blade contact in our implementation, so it produces no complete rollout.
The \emph{Simplified-separation} baseline of Table~\ref{tab:policy-ablation}
remains trainable because it preserves particles and momentum with persistent
CPIC labels. Figure~\ref{fig:kill-revised} shows the difference: ours gives a
stable RL-policy rollout, whereas the Kill-based closed loop diverges at
contact. For a direct visual comparison, we replay the successful trajectory
open-loop with a welded tool on tofu and apply the Kill mask to that same
stable MPM trajectory.

\subsubsection{RL Backbone Ablations}

\begin{figure}[t]
\centering
\begin{tikzpicture}[font=\footnotesize,thick]
\draw[->] (0,0) -- (5.2,0) node[right] {\scriptsize env steps ($\times10^4$)};
\draw[->] (0,0) -- (0,3.0) node[above] {\scriptsize episode return};
\foreach \x/\lab in {1/2,2/4,3/6,4/8,5/10}
  \draw (\x,0) -- (\x,-0.06) node[below] {\scriptsize \lab};
\foreach \y/\lab in {0.5/100,1.0/200,1.5/300,2.0/400,2.5/500}
  \draw (-0.05,\y) -- (0,\y) node[left,xshift=-2pt] {\scriptsize \lab};
\draw[red!70!black,dashed,smooth,domain=0:5,samples=80]
  plot (\x,{0.10+0.20*sin(220*\x)*exp(-0.9*\x)});
\node[red!70!black,anchor=west] at (5.05,0.28) {\ppo{} (1 env)};
\draw[gray!70!black,dotted,thick] (0,1.625) -- (5.2,1.625);
\node[gray!60!black,anchor=south,font=\scriptsize] at (2.0,1.66)
  {scripted demo ($\sim+325$)};
\draw[blue!70!black,smooth,domain=0:5,samples=80]
  plot (\x,{0.40+1.24*(1-exp(-1.6*\x))+0.03*sin(220*\x)*exp(-0.5*\x)});
\node[blue!70!black,anchor=south west] at (4.2,1.72) {\ppo{} (64 env)};
\draw[teal!70!black,smooth,domain=0:5,samples=80]
  plot (\x,{1.625-0.235*(1-exp(-2.5*\x))+0.03*sin(260*\x)*exp(-0.6*\x)});
\node[teal!70!black,anchor=north west] at (4.25,1.34) {\ppo{}+BC};
\end{tikzpicture}
\caption{Observed training behavior on the main \texttt{cut\_tofu} task:
single-environment \ppo{} from random initialization, vectorized \ppo{},
and BC-warm-started \ppo{}. The curves summarize the original
training observations and are not used for statistical comparison; the
controlled multi-seed result is Table~\ref{tab:policy-ablation}.}
\label{fig:ppo-curves-revised}
\end{figure}
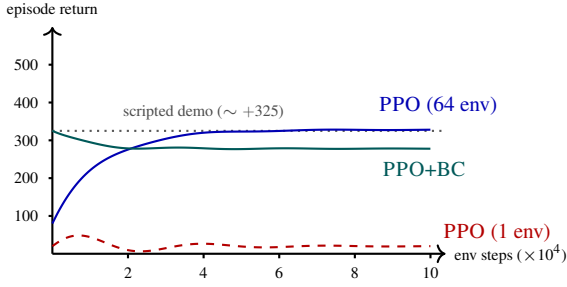

In our main cutting tasks, successful trajectories are difficult to discover through random exploration. On
\texttt{cut\_tofu}, single-environment \ppo{} from random initialization
collapses near the do-nothing return, while a 64-environment vectorized
run reaches approximately $+328$ mean return and a BC-warm-started
single-environment policy retains approximately $+278$. These observations
motivate the curriculum of Sec.~\ref{sec:implementation}; they are distinct
from the matched three-seed damage-model comparison in
Table~\ref{tab:policy-ablation}. Figure~\ref{fig:ppo-curves-revised}
summarizes these three training behaviors on the main
\texttt{cut\_tofu} task.

\begin{figure}[t]
\centering
\begin{minipage}{0.31\columnwidth}\centering
\includegraphics[width=\textwidth]{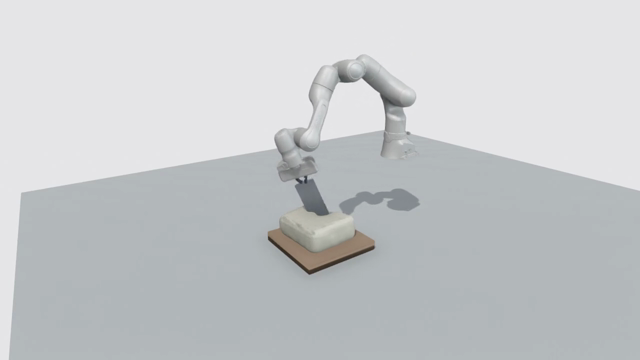}\\
{\scriptsize $t=0.5\,\mathrm{s}$}
\end{minipage}\hfill
\begin{minipage}{0.31\columnwidth}\centering
\includegraphics[width=\textwidth]{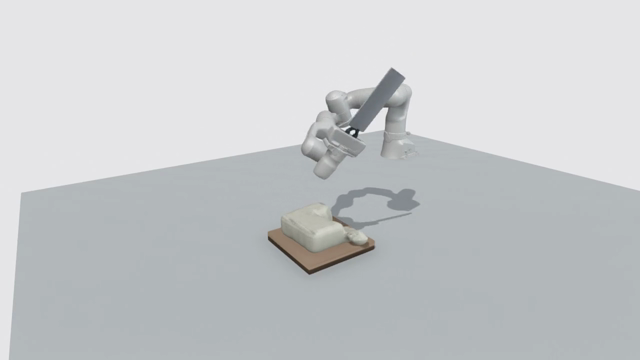}\\
{\scriptsize $t=2.0\,\mathrm{s}$}
\end{minipage}\hfill
\begin{minipage}{0.31\columnwidth}\centering
\includegraphics[width=\textwidth]{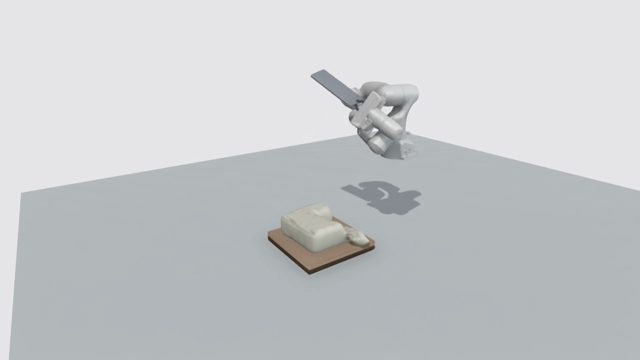}\\
{\scriptsize $t=3.5\,\mathrm{s}$}
\end{minipage}
\caption{Representative \sac{} rollout on \texttt{cut\_tofu}. The actor
lifts and tilts the blade away from the material, leaving the block
undisturbed.}
\label{fig:sac-fails-revised}
\end{figure}

We also report the \sac{} ablation to assess whether changing the RL algorithm addresses the exploration difficulty. Across four attempted
variants, the entropy-regularized actor moves the blade away from the
narrow trajectory of descent, engagement, and cutting and remains at the do-nothing
floor. This negative result does not establish a general disadvantage of
\sac{}; it records the failure mode under our observation, reward, and
training budget. Figure~\ref{fig:sac-fails-revised} shows a representative
rollout exhibiting this behavior.

\section{Conclusion}

We presented \ours, a Gym-like \mpm{} simulator
for RL-based dexterous manipulation of multi-state materials. Its primary
algorithmic contribution couples stress-driven phase-field damage to a
tool-centered localized solve on a uniform \mpm{} background grid and
uses residual-controlled MGPCG to accelerate that local system. Combining it with
two-way rigid-tool coupling on an FR3-arm Gym interface in which
the tool is held by the gripper rather than welded to the
end-effector, \ours covers soft solids, brittle materials, granular
media, and viscoplastic media within a single environment. We
exercised the simulator on six demonstrations:
RL-trained \ppo{} agents (vectorized and BC-warm-started variants)
on two cutting tasks that bracket the stiffness/toughness axis
(\texttt{cut\_tofu}, \texttt{cut\_peanut\_brittle}), plus scripted
simulation-coverage scenes for two intermediate cutting demos
(\texttt{cut\_jelly}, \texttt{cut\_wax}) and the viscoplastic
(\texttt{scoop\_icecream}) and granular (\texttt{shovel\_sand})
material classes.

\paragraph*{Limitations and future work.}
\label{sec:future}
The dominant limitation of \ours is the wall-clock cost of \mpm{}
itself. Even with MGPCG restricted to $\mathcal{N}(\mathbf{T})$,
a single full-physics rollout costs an order of magnitude more time
than a rigid-body environment, and each end-to-end RL training run
on \texttt{cut\_tofu} or \texttt{cut\_peanut\_brittle} consumes several hours
of single-GPU compute. This is why we focused the RL evaluation of
this paper on cutting and treated the viscoplastic
(\texttt{scoop\_icecream}) and granular (\texttt{shovel\_sand})
scenes as simulation-coverage demonstrations rather than as RL
benchmarks. Closing this gap is the natural next step: extending
the two-stage pipeline of Sec.~\ref{sec:implementation} with broader
proxy distillation, parallel MGPCG batching across patches, or
amortized damage solves should make multi-task RL training
tractable
on a single workstation. Beyond compute, brittle damage parameters
are currently tuned per material class rather than learned, friction
uses a single Coulomb $\mu$ per material, the tool-tracking patch
assumes a single primary gripper, and transfer to a physical robot
remains to be evaluated.

\section*{Acknowledgements}
We thank Libin Liu, Siyuan Luo, and Ziqiu Zeng for helpful discussions
and insights.

\clearpage

\onecolumn
\begin{multicols}{2}
\bibliographystyle{eg-alpha-doi}
\bibliography{references}
\end{multicols}

\end{document}